\documentclass[reprint,prl,aps,nobalancelastpage]{revtex4-1}
\usepackage{graphicx}
\usepackage{color}
\usepackage{dcolumn}
\usepackage{bm}
\usepackage{amssymb}
\usepackage{color,amsmath}
\usepackage{soul,xcolor}
\setstcolor{red}
\usepackage{epstopdf}

\begin{document}

\title{Nonlinear twistoptics at symmetry-broken interfaces}

\author{Kaiyuan Yao$^{1*}$}
\author{Nathan R. Finney$^{1*}$} 
\author{Jin Zhang$^{2}$}
\author{Samuel L. Moore$^{3}$}
\author{Lede Xian$^{2}$}
\author{Nicolas Tancogne-Dejean$^{2}$}
\author{Fang Liu$^{4}$}
\author{Jenny Ardelean$^{1}$}
\author{Xinyi Xu$^{1}$}
\author{Dorri Halbertal$^{3}$}
\author{K. Watanabe$^{5}$} 
\author{T. Taniguchi$^{6}$}
\author{Hector Ochoa$^{3}$}
\author{Ana Asenjo-Garcia$^{3}$}
\author{Xiaoyang Zhu$^{4}$}
\author{D. N. Basov$^{3}$}
\author{Angel Rubio$^{2,7}$}
\author{Cory R. Dean$^{3}$}
\author{James Hone$^{1,\dagger}$}  
\author{P. James Schuck$^{1,\dagger}$}

\affiliation{$^{1}$Department of Mechanical Engineering, Columbia University, New York, NY, USA}
\affiliation{$^{2}$Max Planck Institute for the Structure and Dynamics of Matter, Luruper Chaussee 149, 22761 Hamburg, Germany}
\affiliation{$^{3}$Department of Physics, Columbia University, New York, NY, USA}
\affiliation{$^{4}$Department of Chemistry, Columbia University, New York, NY, USA}
\affiliation{$^{5}$Research Center for Functional Materials, National Institute for Materials Science, 1-1 Namiki, Tsukuba 305-0044, Japan}
\affiliation{$^{6}$International Center for Materials Nanoarchitectonics, National Institute for Materials Science, 1-1 Namiki, Tsukuba 305-0044, Japan}
\affiliation{$^{7}$Center for Computational Quantum Physics, Simons Foundation Flatiron Institute, New York, NY 10010 USA}
\affiliation{$^{*}$These authors contributed equally to this work.}
\affiliation{$^{\dagger}$ jh2228@columbia.edu (J.H.); pjs2191@columbia.edu (P.J.S.)}
\maketitle

\textbf{Broken symmetries induce strong nonlinear optical responses in materials and at interfaces. Twist angle can give complete control over the presence or lack of inversion symmetry at a crystal interface, and is thus an appealing knob for tuning nonlinear optical systems. In contrast to conventional nonlinear crystals with rigid lattices, the weak interlayer coupling in van der Waals (vdW) heterostructures allows for arbitrary selection of twist angle, making nanomechanical manipulation of fundamental interfacial symmetry possible within a single device~\cite{Finney2019, Ribeiro-Palau2018}. Here we report highly tunable second harmonic generation (SHG) from nanomechanically rotatable stacks of bulk hexagonal boron nitride (BN) crystals, and introduce the term twistoptics to describe studies of optical properties in dynamically twistable vdW systems. We observe SHG intensity modulated by a factor of more than 50, polarization patterns determined by moir\'e interface symmetry, and enhanced conversion efficiency for bulk crystals by stacking multiple pieces of BN joined by symmetry-broken interfaces. Our study provides a foundation for compact twistoptics architectures aimed at efficient, scalable, and tunable frequency-conversion, and demonstrates SHG as a robust probe of buried vdW interfaces.}


Optical frequency conversion is a nonlinear process where an electromagnetic (EM) field strongly interacts with materials to generate light at new colors. Processes such as second harmonic generation (SHG), optical rectification, and optical parametric amplification arise exclusively from non-centrosymmetric crystals in the limit where EM fields vary slowly over critical length scales of the system ({\it i.e.} the electric-dipole limit)~\cite{Shen1984,Heinz1991}. For state-of-the-art tunable light sources, optical information processing, and quantum optics~\cite{Boyd2003,Garmire2003,Chang2014, Fiorentino2007}, a compact solid-state platform with a dynamically tunable and efficient nonlinear response is highly desirable. Despite recent progress~\cite{Cai2011,Seyler2015,Chen2019}, this remains a challenge.

In this letter we provide a first-time demonstration of highly tunable nonlinear twistoptics in homostructures of hexagonal boron nitride (BN) by combining twistronic~\cite{Carr2017} techniques for assembly and control~\cite{Finney2019, Ribeiro-Palau2018, Kim2016} with powerful nonlinear optical probes. We investigate twist angle, layer number and its parity, and pump field wavelength as tuning parameters for second-order nonlinear optical response in these all-BN devices. In contrast to prior works observing SHG from stationary vdW heterostructures \cite{kim2013, Hsu2014, Yao2020, Liu903}, our study highlights a unique dynamical modality, in which a deeply buried two-dimensional interface between three-dimensional bulk BN crystals is nanomechanically twisted for drastic nonlinear optical response tuning. We also demonstrate strategies to greatly enhance SHG intensity by stacking multiple bulk crystals with symmetry-broken interfaces. The results pave the way for future nano-optomechanical devices based on vdW heterostructures.

\begin{figure*}[ht]
\includegraphics[width=7.0 in]{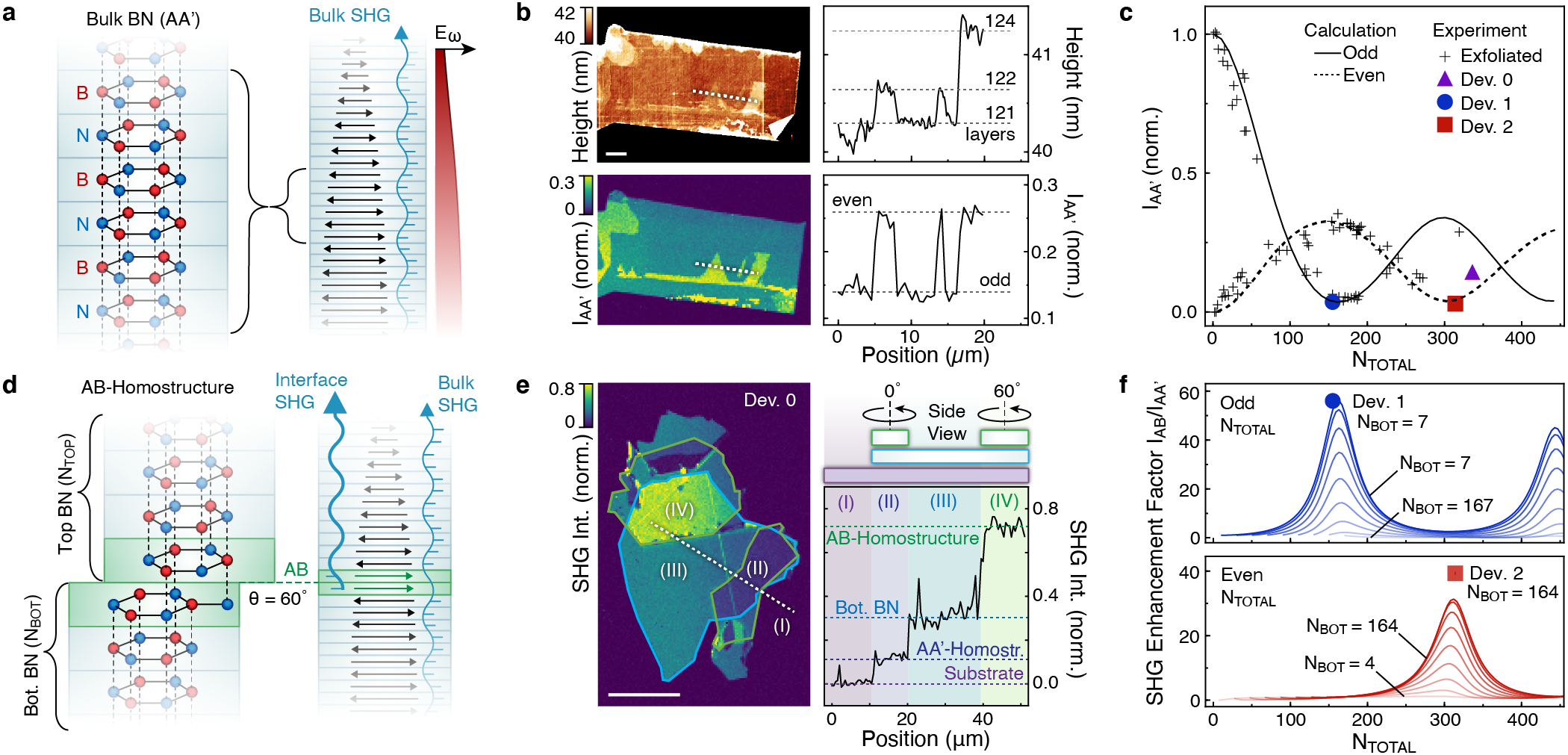}
\caption{\textbf{SHG from bulk BN crystals and homostructures.}
\textbf{a}, Schematics of crystal structure and bulk SHG (wavy blue arrow) for BN crystal with AA' stacked layers. Red B (blue N) labels boron (nitrogen) atoms. Black arrows represent nonlinear dipole moments within each layer. $\boldsymbol E_{\omega}$ denotes pump field.
\textbf{b}, AFM topography (top) and confocal SHG imaging (bottom) of a bulk BN crystal, with corresponding linecuts (right) along the white dashed lines. Scale bar is 5 $\mu$m.
I$_{AA'}$ denotes SHG intensity for AA'-stacked BN normalized to that from monolayer BN.
\textbf{c}, I$_{AA'}$ vs. total layer number $N_{TOTAL}$ for AA'-stacked BN. Solid (dashed) lines are nonlinear transfer matrix calculation results for BN with an odd (even) $N_{TOTAL}$. Grey plus-signs (colored markers) represent data for exfoliated BN (AA'-homostructures).
\textbf{d}, Schematic of a BN homostructure with an AB-stacked interface formed by twisting top BN to $\theta=60^{\circ}$. Strong dipole-allowed SHG (thick wavy blue arrow) arises from the interfacial bilayer with broken inversion symmetry (green highlight). N$_{TOP}$ (N$_{BOT}$) denotes the layer number in top (bottom) BN.
\textbf{e}, SHG image of a BN homostructure. Scale bar is 20 $\mu$m.
Top-right schematic shows top BN (outlined green), comprising two pieces twisted 0$^{\circ}$ (region II) and 60$^{\circ}$ (region IV) with respect to bottom BN (outlined blue), all atop a fused silica substrate (outlined purple). SHG intensity along the white dashed line is shown on bottom right, with dashed lines representing calculation results.
\textbf{f}, SHG enhancement factor I$_{AB}$/I$_{AA'}$ as a function of $N_{TOTAL}$, where I$_{AB}$ is total SHG intensity for AB-homostructures. Solid lines are calculation results. Blue (red) dot (square) are experimental data for Device 1 (2). Cascading curves separated by 20 layer increments.}
\label{fig:1}
\end{figure*}

Bulk BN crystals follow an AA’ interlayer stacking sequence~\cite{Pease1950} (Fig.\ref{fig:1}a). Thus a given BN crystal belongs to the centrosymmetric D$_{3d}$ point group if the total number of layers N is even, while BN crystal with odd N belongs to the non-centrosymmetric D$_{3h}$ point group. The second-order nonlinear polarization $\boldsymbol P_{2\omega}$ from BN can be expressed as a multipole expansion~\cite{Shen1984, Heinz1991}, $\boldsymbol P_{2\omega}=\boldsymbol \chi^{(2)}:\boldsymbol E_{\omega} \boldsymbol E_{\omega} + \boldsymbol \chi_q^{(2)} : \boldsymbol E_{\omega} \nabla \boldsymbol E_{\omega}+ \cdots$, where $\boldsymbol \chi^{(2)}$ and $\boldsymbol \chi_q^{(2)}$ are the second-order susceptibility tensors for dipole and quadrupole moments, respectively, and $\boldsymbol E_{\omega}$ is the pump electric field. In few-layer BN, SHG is observed only in samples with odd N, where broken centrosymmetry results in an allowed dipole moment per unit volume~\cite{Li2013}. When the BN thickness becomes non-negligible compared to the optical wavelength, thin film interference results in a gradient of the pump field $\boldsymbol E_{\omega}$ that breaks net inversion symmetry in the out-of-plane direction~\cite{Heinz1991,Stepanov2020, Kim2019} (Fig.~\ref{fig:1}a), such that the quadrupole contribution becomes important. Fig.~\ref{fig:1}b shows an exfoliated BN crystal with N ranging from 121 to 124 layers, as determined by combined atomic force microscope (AFM) and SHG data (see methods). In this sample, the region with nominally even N shows bright SHG despite being centrosymmetric, reflecting the quadrupole effect. Meanwhile in noncentrosymmetric regions with odd N, the cancellation between dipole and quadrupole moments leads to weaker SHG owing to an approximate $\pi$ phase shift (See Fig.~\ref{fig:S_SHG_Ratio_Simulation}).

The measured SHG from several BN crystals is summarized in Fig.\ref{fig:1}c with pump wavelength at 860 nm. The experimental data clearly fall onto two branches, consistent with the dipolar signal's layer number parity dependence. The data are well described by nonlinear transfer matrix simulations for odd and even layer number N~\cite{Bethune1989} that incorporate the nonlocal multipolar effects (Supplementary Section ~\ref{SI_nlTM}). The evolution and reversal of even-odd SHG contrast results from the variation of quadrupole contribution with N, allowing nonlinear response tailoring by layer number engineering. With the layer number parity and SHG linked, we can corroborate layer number estimates for bulk BN as in Fig.~\ref{fig:1}b. The data in Fig.~\ref{fig:1}c also includes AA'-homostructures made by stacking two exfoliated BN crystals, confirming that the AA’ interface does not give rise to additional SHG.

\begin{figure*}[ht]
\includegraphics[width=7.0 in]{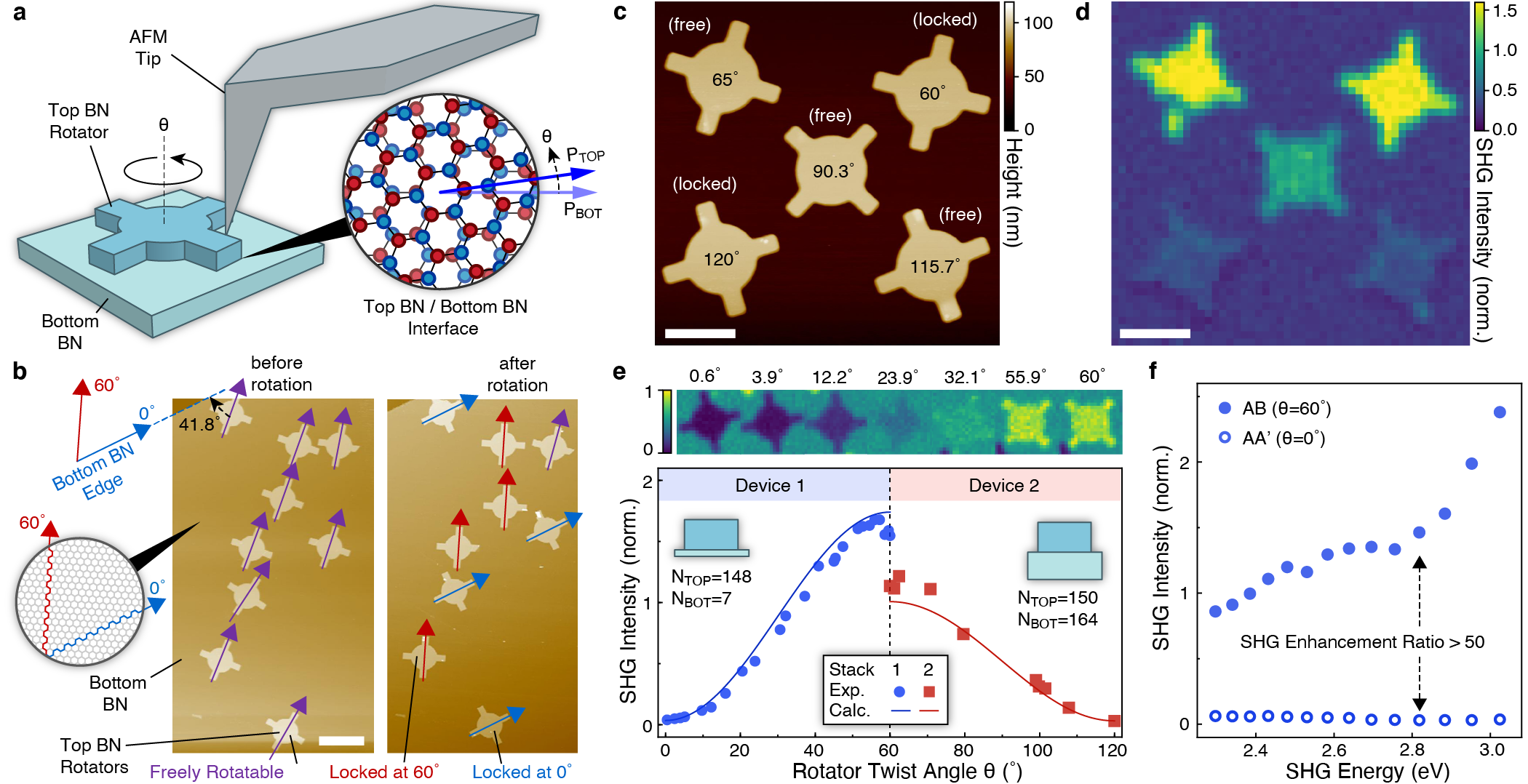}
\caption{\textbf{Dynamical tuning of SHG from BN micro-rotators.}
\textbf{a} Schematic of a top BN rotator pushed by an AFM on bottom BN. Inset illustrates a top view of the interface at twist angle $\theta$. Blue arrows represent tunable nonlinear dipole moments.
\textbf{b} AFM images of BN rotators dialed to locking positions at 0$^{\circ}$ (blue arrows) and 60$^{\circ}$ (red arrows) with respect to bottom BN. Bottom BN edge (blue dashed line) is parallel to a crystallographic direction of the bottom BN (illustration, left). Purple arrows indicate orientations of freely rotatable top BN, one angle measurement shown as 41.6$^{\circ}$. Scale bar is 5 $\mu$m. 
\textbf{c} AFM image and \textbf{d} SHG image of BN rotators dialed to different twist angles. Scale bars in \textbf{c-d} are 3 $\mu$m. 
\textbf{e} SHG intensity vs. $\theta$ data for Device 1 (blue dots) and Device 2 (red squares), with with calculated trends (solid lines), and layer numbers indicated by insets. SHG images for rotation states $\theta<60^{\circ}$ shown above.
\textbf{f} Energy dependent SHG intensity for AB ($\theta=60^{\circ}$, solid markers) and AA’ ($\theta=0^{\circ}$, open markers) interfacial configurations. Data in \textbf{d-f} normalized to monolayer BN SHG intensity at 2.88 eV.}
\label{fig:2}
\end{figure*}

In a BN-BN homostructure, local symmetry can be broken at the interface by introducing a twist angle $\theta$ between the top and bottom crystals. Fig.~\ref{fig:1}d depicts the case of $\theta = 60^{\circ}$, for which AB interfacial stacking is the lowest energy configuration based on first-principles calculations (Fig.~\ref{fig:S_DFT}). In AB-stacked BN, boron (nitrogen) atoms in the top BN layer are located right above the nitrogen atoms (hollow sites) of the adjacent BN layer, to create a non-centrosymmetric interfacial bilayer.  Strong dipole-allowed SHG arises from this interface, similar to that from AB-stacked bilayer BN~\cite{kim2013}, in coherent superposition with bulk SHG engineered by layer number N. Fig.~\ref{fig:1}e shows an SHG image of a BN homostructure created by breaking a single N-layer flake, rotating selected pieces by 60$^{\circ}$, and subsequently stacking to create 2N-layer AA'- and AB-homostructures (regions II and IV, respectively), as well as the N-layer source crystal (region III).Prior to stacking, the BN was verified to have constant layer number over the regions of interest. The AB-homostructure shows enhanced SHG due to symmetry breaking, while the AA’-homostructure shows significantly weaker signal. To evaluate SHG tunability, we evaluate the SHG enhancement factor $I_{AB}$/I$_{AA'}$, where $I_{AB}$ and $I_{AA'}$ are defined as the SHG from the AB- and AA'-homostructure configurations of a given single device. As shown by Fig.~\ref{fig:1}f, the enhancement factor can be optimized by careful selection of layer number for both the top and bottom BN crystals, which controls the magnitude and phase of bulk and interface SHG. We find that optimum designs can be specifically engineered with large $I_{AB}$ comparable to monolayers and small $I_{AA'}$ close to zero, as in Devices 1 and 2 that feature a dominant SHG contribution from the interface, with minimum SHG contribution from the bulk (Fig.~\ref{fig:S_SHG_Ratio_Simulation} and Supplementary Section~\ref{SI_Gain_Factor_Optimization}).

\begin{figure*}[ht]
\includegraphics[width=7.0 in]{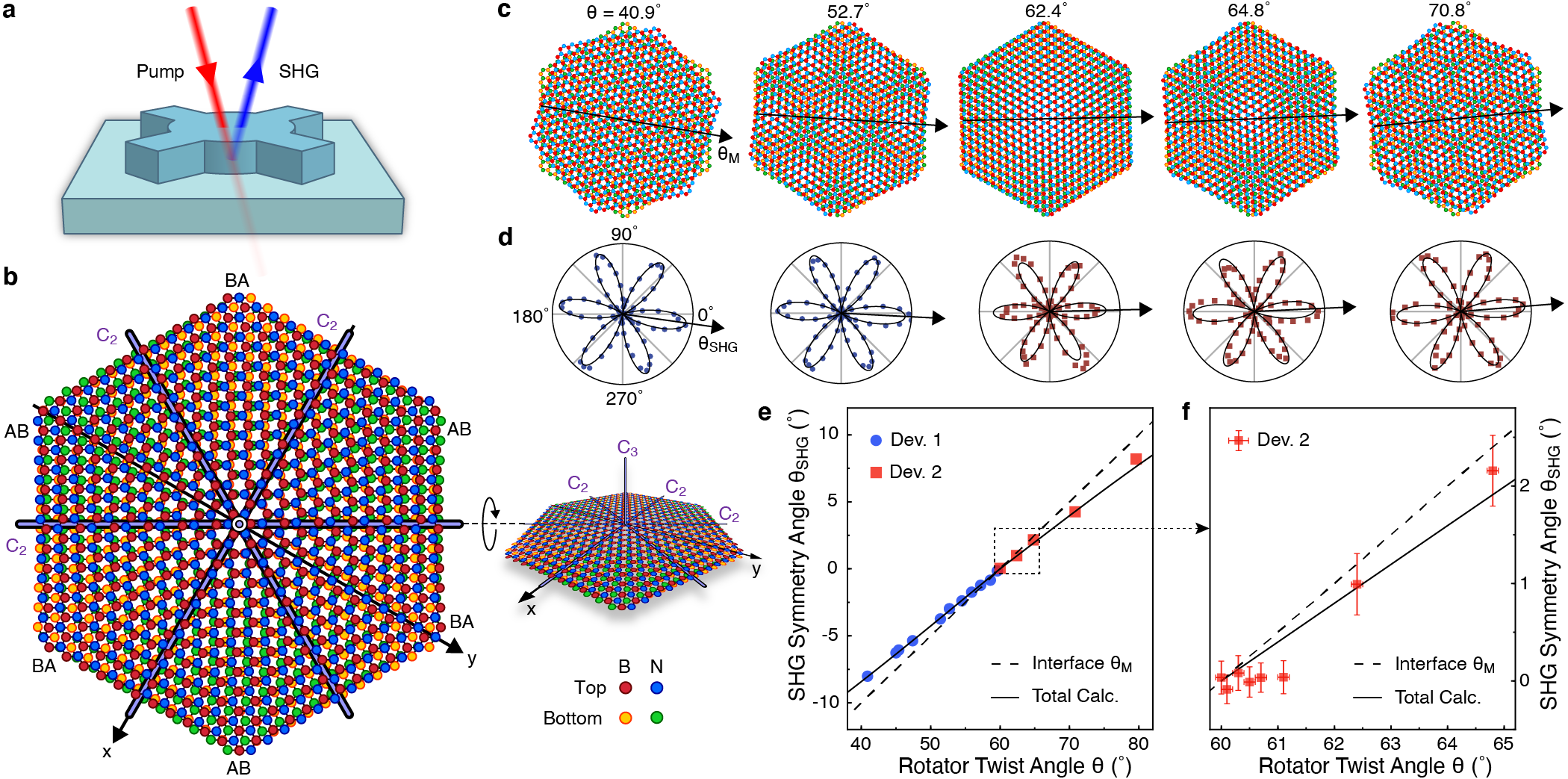}
\caption{\textbf{Probing tunable crystal symmetry by polarization resolved SHG.}
\textbf{a} Schematic of SHG as a selective probe to a symmetry-broken interface.
\textbf{b} Moir\'e unit cell formed by the interfacial bilayer at $\theta=63^{\circ}$, with rotational invariance along three in-plane C$_2$ axes as labeled (left, right), and an out-of-plane C$_3$ axis (right).
\textbf{c} Interface moir\'e patterns and \textbf{d}, corresponding measured SHG polarization patterns at various twist angles $\theta$. Arrows $\theta_M$ in \textbf{c} indicate one of the three C$_2$ axes of the moir\'e pattern as shown by \textbf{b}. Blue(red) dots(squares) are measured from Devices 1(2). Arrows $\theta_{SHG}$ indicate the SHG symmetry angle determined from the fit of the polarization data displayed (solid line).
\textbf{e} Measured $\theta_{SHG}$ as a function of rotator twist angle $\theta$. Dashed line follows $\theta_{SHG}=\theta_M$ as the expected trend for the interfacial bilayer. Solid line shows nonlinear transfer matrix calculation results involving the entire bulk structure.
\textbf{f} Device 2 data near 60$^{\circ}$ (black dashed box in \textbf{e}).}

\label{fig:3}
\end{figure*}

With an understanding of the SHG response from bulk BN and the buried symmetry broken interface in BN homostructures, we can design nano-mechanically rotatable all-BN devices to considerably modulate the SHG response. We tune the interface configuration continuously between the two extremes of AA' and AB stacking by adjusting the twist angle with an AFM tip~\cite{Chari2016, Ribeiro-Palau2018, Finney2019}. Fig.~\ref{fig:2}a illustrates nanomechanical rotation of a BN homostructure to a twist angle of $\theta$, which is defined as a counterclockwise (CCW, as seen from the top) rotation from the AA' configuration where $\theta=0^{\circ}$. Fig.~\ref{fig:2}b shows AFM scans of an assembled rotatable homostructure, comprising an array of bulk BN micro-rotators atop a larger BN substrate. These can be rotated such that some become crystallographically aligned at extremes that are $60^{\circ}$ apart. At these extremes the rotators become locked, and can no longer be rotated by pushing (Fig.~\ref{fig:S_locking}). As shown by Fig.~\ref{fig:2}c-d, SHG imaging can further reveal whether the locking angles correspond to AA' or AB interfacial stacking. Depending on the initial placement of the BN rotators during assembly, one can either operate in the $\theta=0^{\circ}$ to $60^{\circ}$ range where CCW rotation transitions a given rotator from AA' to AB interface as in Device 1, or the $\theta=60^{\circ}$ to $120^{\circ}$ range where CCW rotation transitions a given rotator from AB to AA', as in Device 2. The uniqueness of twist configuration over a $120^{\circ}$ interval that respects the handedness of the twist angle definition is consistent with the three-fold rotational symmetry of the bulk BN lattice.

Fig.~\ref{fig:2}e shows the tuning of SHG intensity with twist angle $\theta$ for two devices, each with different thicknesses and therefore different ranges in SHG intensity modulation. Despite large differences in their bulk thicknesses, both Device 1 and 2 exhibit behavior consistent with the changing interface symmetry---highest (lowest) SHG is observed at close to AB (AA') interface configuration. Importantly, an SHG modulation enhancement greater than 50 can be achieved between AB and AA' interface configurations. For comparison, this is about ten times larger than achievable by electrical tuning mechanisms based on third-order susceptibilities in plasmonic cavities~\cite{Cai2011}, and excitonic SHG from tunable oscillator strength in monolayer semiconductors~\cite{Seyler2015, Yao2017}. Moreover, the high tunability of SHG intensity persists over a broad band of frequencies in the visible spectrum, covering at least several hundred meV as shown by Fig.~\ref{fig:2}f. The operation frequency window is more than an order of magnitude wider than in devices relying on plasmonic or excitonic resonances~\cite{Cai2011,Seyler2015}. Furthermore, the SHG intensity for an AB-interface increases substantially at higher energy, entirely consistent with a predicted exciton-influenced smaller optical band gap as compared to the native AA' stacked BN crystal~\cite{ochoa2020, Xian2019, Wirtz2006, Wirtz2009}.

\begin{figure*}[ht]
\includegraphics[width=7.0 in]{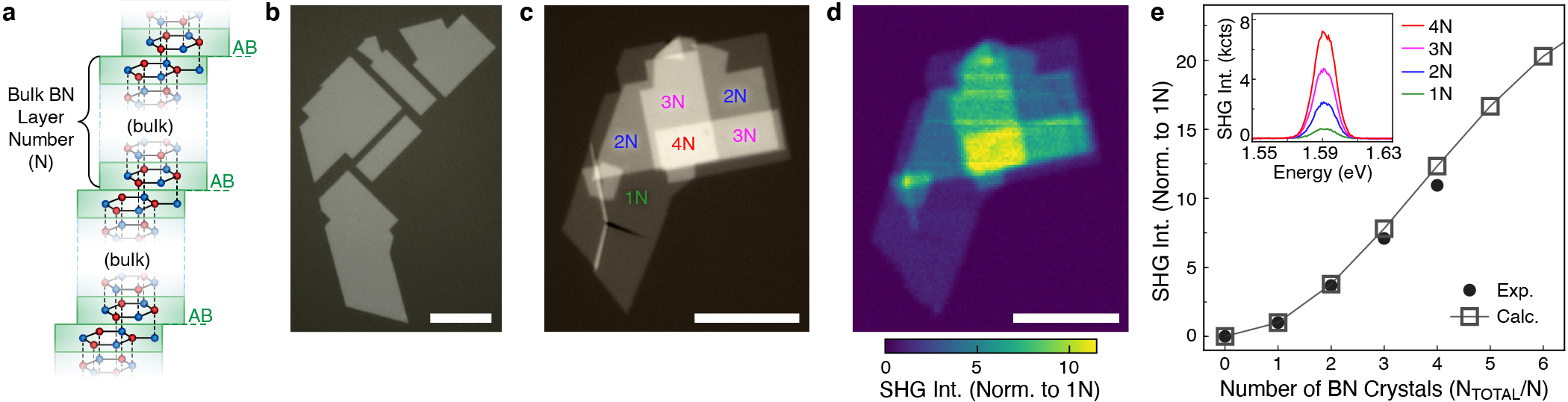}
\caption{\textbf{SHG intensity enhancement by coupling multiple bulk BN crystals.}
\textbf{a}, Schematic for a stack of aligned N-layered bulk BN crystals separated by AB interfaces.
\textbf{b}, Optical image of lithographically patterned BN crystal on fused silica.
\textbf{c}, Optical and \textbf{d}, confocal SHG image of a  device assembled from patterned BN shown in \textbf{b}, featuring regions with 1N-4N layer thicknesses (N = 25), indicated with colored labels. Scale bars for \textbf{b-d} are all 10 $\mu$m.
\textbf{e}, Scaling of SHG intensity as a function of the number of aligned bulk BN crystals in the stack. Data are normalized to intensity from the 1N layer region. Solid circle markers are experimental data and open square markers are calculation results. Inset shows SHG emission spectra from regions of different thickness.}
\label{fig:4}
\end{figure*}

Nonlinear optics can serve as an accurate and selective probe of deeply buried interfaces~\cite{Stepanov2020, Shen2006}. For BN rotators with minimized bulk SHG, the infrared pump beam penetrates the entire structure and the reflected SHG reports information from the symmetry-broken interface, as illustrated by Fig.~\ref{fig:3}a. Moreover, the polarization patterns of SHG directly reflect the point group of the lattice~\cite{Heinz1991,Shen1984}, and have been relied upon to understand properties of moir\'e superlattices in vdW heterostructures~\cite{bai2019}. We can therefore probe the tunable moir\'e symmetry at the deeply buried vdW interface by examining polarization-resolved SHG from the micro-rotators. Fig.~\ref{fig:3}b illustrates the unit cell of the moir\'e pattern formed by the interfacial bilayer BN near AB stacking, which belongs to the non-centrosymmetric D$_3$ point group. With broken inversion symmetry, the electric dipole moment plays the dominant role in SHG from the bilayer interface. The non-vanishing elements of the $\boldsymbol\chi^{(2)}$ tensor responsible for in-plane polarization are $-\chi^{(2)}_{xxx}=\chi^{(2)}_{xyy}=\chi^{(2)}_{yxy}=\chi^{(2)}_{yyx}$ \cite{Shen1984}. In our measurements where pump and SHG collection are collinearly polarized, a sixfold polarization pattern is expected since the SHG intensity varies as $I_{SHG \parallel}\propto cos^2(3(\theta_{POL}-\theta_M))$, where $\theta_{POL}$ is the light polarization angle in the lab frame and $\theta_M$ is the moir\'e angle defined here as orientation of the three equivalent C$_2$ axes (Fig.~\ref{fig:3}b). The interfacial moir\'e patterns and experimental SHG polarization patterns from micro-rotators are compared in Fig.~\ref{fig:3}c-d over a broad range of twist angle $\theta$. The above equation is fit to the polarization data measured from devices 1 and 2. The measured SHG symmetry angle $\theta_{SHG}$ is found to closely follow the expected moir\'e angle that evolves as $\theta/2$.

Far from $\theta=60^{\circ}$, a deviation is observed between $\theta_{SHG}$ and $\theta_M$, as shown by Fig.~\ref{fig:3}e. The difference is attributed to finite residual bulk SHG, which can be included in the transfer matrix calculation that accounts for the entire structure. However, at small twist angle, $\theta$ = 60$^{\circ}$ to 61.2$^{\circ}$, the SHG symmetry angle is found to remain at a constant value as shown by the zoom-in plot of Fig.~\ref{fig:3}f. The bilayer moir\'e model and the transfer matrix model shown here both assume a rigid lattice for the BN, and neither successfully predicts the pinning of polarization symmetry in this range. Lattice relaxations and soliton formation have been observed in buried mismatched planes within bulk BN crystals~\cite{Ni2019}. Furthermore, calculation of the stacking energy for bilayer BN twisted near 60$^\circ$ reveals a strong onset of mechanical relaxation as the twist angle decreases past 61$^\circ$ (~\ref{fig:S_relax}). Therefore we hypothesize that interfacial atomic reconstruction may be providing the requisite conditions for this effect.

The results discussed thus far reveal highly tunable SHG to be an effective diagnostic interrogator of deeply buried vdW interfaces, and illustrates nanomechanical rotation as a non-invasive remote control knob for interfacial nonlinearities. Furthermore, these results point to strategies for further enhancing the efficiency of vdW nonlinear optical devices towards practical applications, such as compact optical parametric oscillators, high harmonic generation, and spontaneous down-converters for entangled photon pairs~\cite{Garmire2003, Chang2014, Fiorentino2007,Trovatello2019}. As shown by Fig.~\ref{fig:4}a, multiple AB interfaces between BN thin films can be coherently coupled within the phase matching limit to enhance SHG intensity. This is achieved by lithographically cutting a BN crystal with odd layer number N and stacking the pieces with AB interfacial orientation (Supplementary Section ~\ref{SI_Fab_Stationery_Devices}). Fig.~\ref{fig:4}b shows lithographically patterned pieces of a BN crystal with N = 25, and Fig.~\ref{fig:4}c shows the stacked sample with regions of different numbers of bulk BN crystals layered on top of one another. Fig.~\ref{fig:4}d shows confocal SHG imaging of the resulting structure shown in Fig.~\ref{fig:4}c. Owing to coherent superposition between neighboring layers, the SHG intensity grows super-linearly with the number of crystals forming AB interfaces (Fig.~\ref{fig:4}d-e). The SHG intensity is increased by more than an order of magnitude with homostructure stacks consisting of four BN crystals, measuring a total of 33 nm in thickness, \textit{i.e.} only about 2$\%$ of the SHG pump wavelength. The trend becomes saturated in thicker devices due to onset of phase-mismatch, which could be overcome by employing a quasi-phase-matched structure~\cite{Boyd2003} using twist and layer number.

In conclusion, nonlinear twistoptics in deeply buried symmetry-broken van der Waals interfaces are demonstrated for highly tunable nonlinear optical applications. In rotatable BN homostructures, the twist angle and the underlying moir\'e interface are dynamically controlled by nanomechanical means, resulting in symmetry tuning of the interfacial lattice, and hence drastically modulated SHG. The enhanced SHG arising from the interface in an optimally designed BN homostructure proves to be a sensitive reporter of twist-angle-dependent crystal and optoelectronic properties at the buried interface. With the rapid development in wafer-scale growth of single crystal BN \cite{Lee2018, Wang2019, Chen2020}, and large area disassembly and reassembly techniques~\cite{Liu903}, we envision that the prototypical demonstrations provided here can be scaled up to further enhance device efficiency, and open new possibilities for compact optical microelectromechanical systems.

\section{Methods}
All BN homostructures are assembled using established dry-transfer techniques with a poly-propylene carbonate (PPC) film on a polydimethyl siloxane (PDMS) stamp~\cite{Wang2013}, using a high-precision rotation stage~\cite{Kim2016}. Rotatable structures are fabricated and then mechanically actuated using techniques and processes that are identical to those described in detail in Refs.~\cite{Ribeiro-Palau2018, Finney2019}, sans graphene processing steps. We note that at no stage do the surfaces that form the interface come into contact with the transfer polymers. In the case of the sample featured in Fig. 4, for which the source BN was lithographically defined using a PMMA mask, the residues were removed using oxygen-argon annealing~\cite{Garcia2012}. Unless otherwise mentioned, all devices and exfoliated samples rest atop a fused silica substrate for study. The bottom BN in Device 4 appearing in Supplemental Section S3 and Fig. S3-S4 was exfoliated using  large-area disassembly techniques~\cite{Liu903}. The crystallographic orientations of the top BN rotator and the bottom BN substrate are predetermined by performing polarization-resolved SHG before the assembly (Fig.~\ref{fig:S_Bulk_Polarization}). Straight edges of bulk BN crystals that are found to be typically aligned with either the armchair or zigzag orientation serve as references for defining the straight-arms of the rotators.

SHG measurements performed are generically similar to those in Ref.~\cite{Yao2020}. A Ti:Saph femptosecond pulsed laser and optical parametric oscillator are used as pump source, with typical power on the order of a few mW. An objective with numerical aperture of 0.95 is used for confocal imaging, with a typical beam spot diameter of about 550 nm. For accurate measurement of nonlinear intensity, each sample is always piezo-scanned along the z-direction to optimize focus. An electron-multiplying CCD is used for fast signal collection with spectral resolution. For a complete diagram of the measurement setup see Supplementary Fig.~\ref{fig:S_setup}.

For the bulk BN thickness and layer number parity dependent BN measurements shown in Fig.~\ref{fig:1}b-c, the thickness (t) with corresponding layer number (N = t/0.334) of a given BN crystal is measured from AFM scans. The uncertainty in thickness measurements of a given bulk flake BN is estimated to be $\pm$0.5 nm. Confocal SHG measurements for mono-to-few layer steps on the top surface of the BN crystal (corresponding to measured total thickness $N_{Total}$ = ..., N-2, N-1, N, N+1, N+2,...) are compared with the prediction given by the nonlinear transfer matrix calculations, which allows us to assign the layer number parity for the given region based on our model, and therefore the layer number parity of N.

Angle measurements of BN micro-mechanical rotators are extracted from AFM scans using a digital protractor. The twist angle $\theta$ between a given BN rotator and the bottom BN is assigned based on its angle relative to the locked rotators that are 60$^{\circ}$ apart from one another. Assignment of $\theta$ for the locked rotators is given through a comparison of the measured SHG intensity from each rotator and the calculated prediction for AA' and AB stacked bulk-on-bulk BN with comparable thicknesses. Locking is determined by observing an inability to continue pushing the BN rotator without physically damaging it with the AFM tip. Further details regarding this locking behavior is provided in Supplementary Section~\ref{SI_Locking} and Fig.~\ref{fig:S_locking}.

\section*{acknowledgments}
The authors thank Di Xiao, and Jue Wang for helpful discussions and technical support. This research is mainly supported as part of Programmable Quantum Materials, an Energy Frontier Research Center funded by the U.S. Department of Energy (DOE), Office of Science, Basic Energy Sciences (BES), under award DE-SC0019443. Theory support involving DFT calculations for this project is supported by the European Research Council (ERC-2015-AdG694097), the Cluster of Excellence AIM and the  Max Planck Institute - New York City Center for Non-Equilibrium Quantum Phenomena is acknowledged (A.R.). The Flatiron Institute is a division of the Simons Foundation. N.F. acknowledges support from the Stewardship Science Graduate Fellowship program provided under cooperative agreement number DE-NA0003864. This work was also supported by a grant from the Simons Foundation (579913, DH). K.W. and T.T. acknowledge support from the Elemental Strategy Initiative conducted by the MEXT, Japan, Grant Number JPMXP0112101001, JSPS KAKENHI Grant Numbers JP20H00354 and the CREST(JPMJCR15F3), JST.

\section*{Author contributions}
K.Y. and N.R.F. performed the measurements and analyzed the data. K.Y., J.Z., L.X., N.T.-D., D.H., H.O., A.A.-G., and A.R. performed theory calculations. N.R.F. fabricated all devices. F.L. assisted with sample preparation. S.M., J.A., and X.X. assisted with preliminary experimental efforts. K.W. and T.T. grew the BN crystals. H.O., A.A.-G., X.Y.Z., D.N.B, A.R., C.R.D., J.H., and P.J.S advised on the experiments. The manuscript was written with input from all authors.

\section*{Competing interests}
 The authors declare no competing interests.

\bibliographystyle{naturemag}
\bibliography{references}

\clearpage


\renewcommand{\thefigure}{S\arabic{figure}}
\renewcommand{\thesubsection}{S\arabic{subsection}}
\setcounter{secnumdepth}{2}
\renewcommand{\theequation}{S\arabic{equation}}
\renewcommand{\thetable}{S\arabic{table}}
\setcounter{figure}{0} 
\setcounter{equation}{0}

\section*{Supplementary Information}

Supplementary Sections S1-S9, and supplementary figures S1-S13.

\subsection{Nonlinear transfer matrix method}
\label{SI_nlTM}
A numerical model is developed to compute nonlinear frequency conversions in layered van der Waals (vdW) stacks, which allows for implementation of an arbitrary symmetry group, twist angle and nonlinear efficiency into each composite layer. Nonlocal multipole moments of the nonlinear polarization are incorporated. Phase-matching effects and multilayer reflections are included. The method is based on a transfer matrix formalism initially employed for studying generation and propagation of high harmonic electromagnetic waves in multilayer nonlinear thin films~\cite{Bethune1989}. The model proposed in Ref.~\cite{Bethune1989} calculates third harmonic generation. In this work it is adapted to calculate second harmonic generation (SHG) from multilayered hexagonal boron nitride (BN) rotators, with polarization resolution and arbitrary crystal orientation for each layer. Owing to the nature of transfer matrix formalism, it is also convenient to resolve the SHG contribution from each single layer of BN, allowing us to separately compute interface and bulk effects. Each BN layer within the rotator structure is modeled by linear refractive indexes of bulk BN in the visible and near infrared region~\cite{Lee2019}, and the thickness is assumed to be interlayer spacing, i.e. 0.334 nm. Interlayer couplings are not included, and each layer is assumed to be a rigid plate without any lattice reconstruction or strain effects. The depletion of the pump field due to frequency conversion is also neglected, which is a good approximation for relatively low field experiments. This model may be also applied for calculating other complex vdW heterostructures in the frequency range where interlayer coupling is minimal.

\begin{figure*}[ht]
\includegraphics[width=6.0 in]{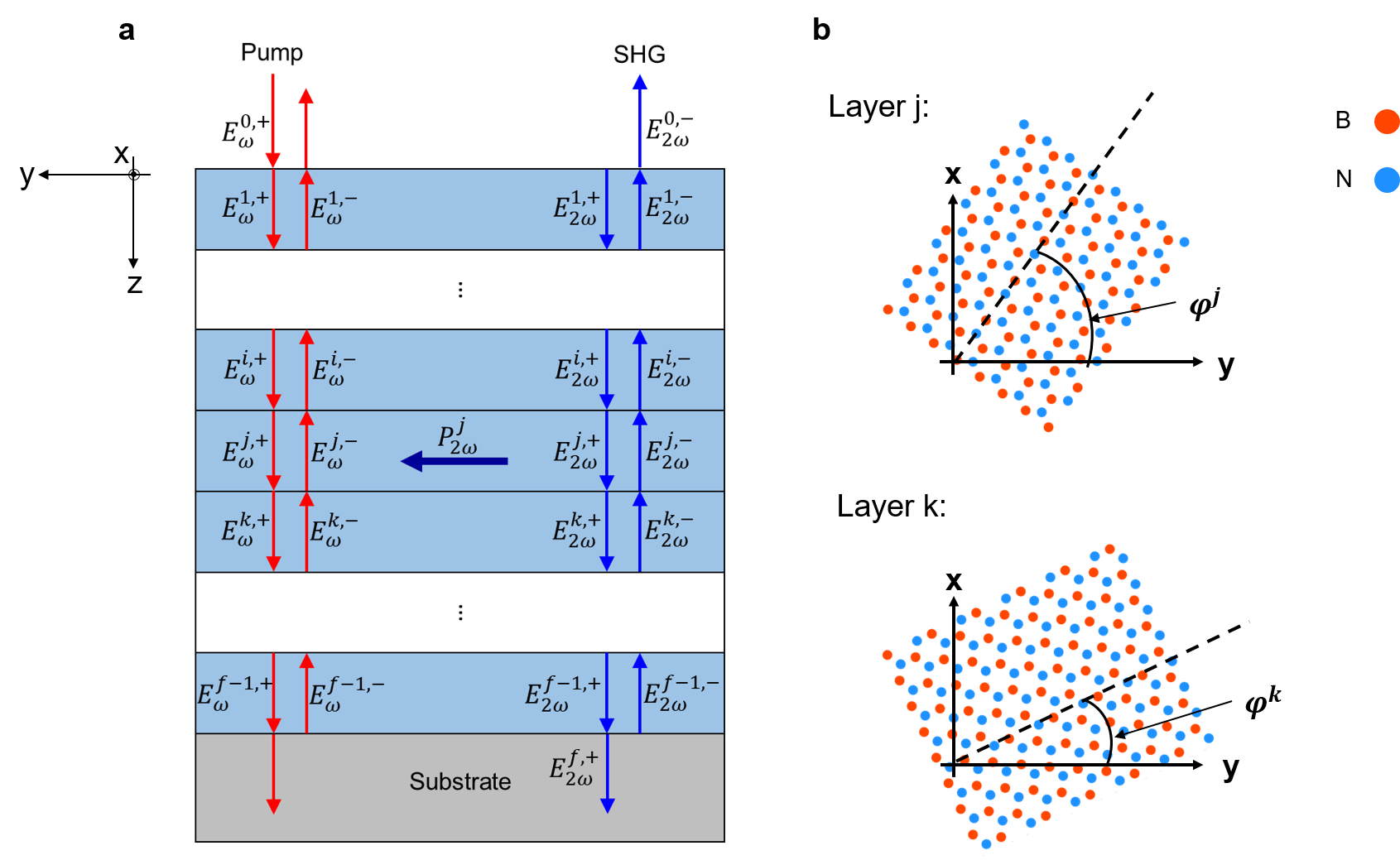}
\caption{\textbf{Schematic of the nonlinear transfer matrix calculation method.}
\textbf{a}, Side-view of multilayered BN on substrate. The forward- and backward-propagating pump field $E^{j,+}_{\omega}$ and $E^{j,-}_{\omega}$ generate nonlinear polarization $P^{j}_{2\omega}$ in layer j, which serves as the source for driving second harmonic field that propagates both forward ($E^{j,+}_{2\omega}$) and backward ($E^{j,-}_{2\omega}$) in the multilayers. The squared norm of reflected SHG field $E^{0,-}_{2\omega}$ corresponds to the experimentally measured quantity. The total SHG response is the coherent sum of SHG field amplitudes driven by nonlinear polarization in each layer. See text for meaning of notations and more detailed model descriptions.
\textbf{b}, Top-view of representative layer $j$ and layer $k$ showing the in-plane coordinate system used for calculation. The angle $\varphi^{j}$ is defined as between the $+y$ direction and the armchair direction of layer $j$. Arbitrary rotation angles can be implemented for each individual layers.
}
\label{fig:S_TMM}
\end{figure*}

The principle of the nonlinear transfer matrix calculation is illustrated by Fig.~\ref{fig:S_TMM} and summarized as follows. In this coordinate system, $+z$ is the out-of-plane direction pointing from air to substrate. Each layer is indexed by numbers $0$, $1$, … $i$, $j$, $k$, … $f-1$, $f$, with $0$ being air, $f$ being the substrate, and $1$, $2$, .. $f-1$ representing monolayer BN. The crystal orientation of each BN layer is described by $\varphi^{j}$ which is the angle between BN armchair direction and the $+y$ direction, as shown by Fig.~\ref{fig:S_TMM}. The pump field in any layer $j$ is described as a two-component pseudovector $\boldsymbol E_{\omega,y}^{j}$ $=[E_{\omega,y}^{j,+}, E_{\omega,y}^{j,-}]$ where $E_{\omega,y}^{j,+}$ ($E_{\omega,y}^{j,-}$) is the complex amplitude of forward (backward) propagating electric field. Without loss of generality the incident pump field is set to be along the $+y$ direction. Since the linear index of BN is in-plane isotropic, the polarization of the pump field in each layer will be preserved and will be along the y direction (but with different phases). For polarization dependent calculations, we rotate crystallographic orientation of the BN layers by setting arbitrary $\varphi^{j}$ for each layer, which is equivalent to a rotation of optical polarization. Fig.~\ref{fig:S_TMM}a illustrates the calculation of SHG contribution from layer $j$. In low pump field regime, depletion of the pump field can be neglected~\cite{Bethune1989}, thus the nonlinear electric field contribution from each layer can be linearly summed together to obtain the total response. In the source layer $j$ with nonzero $\chi^{(2)}$, the nonlinear polarization $\boldsymbol P^{j}_{2\omega,y}$ and $\boldsymbol P^{j}_{2\omega,x}$ generate bound waves at frequency $2\omega$, which are represented by $\boldsymbol E_{S,y}^{j}=[E_{S,y}^{j, +}, E_{S,y}^{j, -}]$ and $\boldsymbol E_{S,x}^{j}=[E_{S,x}^{j, +}, E_{S,x}^{j, -}]$. Then the bound waves in the source layer drives free waves at $2\omega$ frequency (denoted $\boldsymbol E_{2\omega,y}^{j}$and $\boldsymbol E_{2\omega,x}^{j}$) in all the layers, including the reflected SHG wave collected  experimentally in the air medium.

The detailed calculation steps are described below. The wave propagation of the pump field and the SHG field through the multilayer structure are calculated following standard linear optical transfer matrix methods used in previous work ~\cite{Yao2017}. The field pseudovectors in neighouring layers are related by transfer matrices $\boldsymbol M^{ij}$ which are calculated for the proper frequency ($\omega$ or $2\omega$) and polarization ($x$ or $y$) based on the Fresnel reflection and transmission coefficiencts $r^{ij}$ and $t^{ij}$. The total transfer matrix $\boldsymbol T$ for the specific frequency ($\omega$ or $2\omega$) and polarization ($x$ or $y$) can then be calculated as following.

\begin{equation}\label{TMM_1}
\boldsymbol E^i = \boldsymbol M^{ij} \boldsymbol E^i
\end{equation}

\begin{equation}\label{TMM_2}
\boldsymbol M^{ij} = \frac{1}{t^{ij}}   \begin{bmatrix}
                                        1 & r^{ij}\\
                                        r^{ij} & 1
                                        \end{bmatrix}
\end{equation}

\begin{equation}\label{TMM_3}
\boldsymbol T = \boldsymbol M^{f(f-1)} \boldsymbol \Phi^{(f-1)} \boldsymbol M^{(f-1)(f-2)} \boldsymbol \Phi^{(f-2)}...\boldsymbol M^{10}
\end{equation}

\begin{equation}\label{TMM_4}
\boldsymbol \Phi^{i} =    \begin{bmatrix}
                            \phi^i & 0\\
                            0 & \frac{1}{\phi^i}
                            \end{bmatrix}
\end{equation}

Within BN layer $j$, the elementary nonlinear polarization is $P_{2\omega, \alpha}^{j}$ where $\alpha$ denotes the polarization direction ($\alpha$ = x or y). In general, $P_{2\omega, \alpha}^{j}$ includes multipole contributions, with the dipole moment $P_{2\omega, \alpha}^{j,d}$, quadrupole moment $P_{2\omega, \alpha}^{j,q}$, and higher order terms with decreasing magnitudes in the low-field regime:
\begin{equation}\label{TMM_PNL_DQ}
\begin{aligned}
P_{2\omega, \alpha}^{j} & = P_{2\omega, \alpha}^{j,d} + P_{2\omega, \alpha}^{j,q} + ...\\
& =\epsilon_0 \chi^{(2)}_{\alpha\beta\gamma} E_{\omega, \beta}^{j}E_{\omega, \gamma}^{j} + \epsilon_0 \chi^{q,1L}_{\alpha\beta\gamma z} E_{\omega, \beta}^{j} \frac{ \partial E_{\omega, \gamma}^{j}}{\partial z} + ...  
\end{aligned}
\end{equation}
Here $\chi^{(2)}_{\alpha\beta\gamma}$ are elements of the nonlinear susceptibility tensor $\boldsymbol \chi^{(2)}$ with the subscripts $\alpha, \beta, \gamma$ denoting in-plane polarization directions in Cartesian coordinates. $\chi^{q,1L}_{\alpha\beta\gamma z}$ are elements of quadrupolar nonlinear susceptibility tensor $\boldsymbol \chi^{q, 1L}$ for the quadrupole moment from a monolayer BN. The last subscript is fixed to be $z$ as only plane waves are considered in the calculation. $\epsilon_0$ is the vacuum permittivity. $E_{\omega, \beta}^{j}$ is the complex amplitude of the pump electric field in layer $j$ polarized along the $\beta$ direction. Note that we follow the Einstein notation for summing over subscripts $\alpha$, $\beta$, $\gamma$ whenever they are repeated.

Since each monolayer BN belongs to the point group of $D_{3h}$, its nonvanishing $\boldsymbol \chi^{(2)}$ tensor elements are:
\begin{equation}\label{TMM_D3h_Chi2}
\chi^{(2)}_{y'y'y'} = -\chi^{(2)}_{y'x'x'} = -\chi^{(2)}_{x'x'y'} = -\chi^{(2)}_{x'y'x'}  = |\chi^{(2)}|
\end{equation}
with $x'$ and $y'$ being the zigzag and armchair orientations as shown by Fig.~\ref{fig:S_TMM}.

The monolayer quadrupole moments $P_{2\omega, \alpha}^{j,q}$ in ~\eqref{TMM_PNL_DQ} will vanish due to the reflection symmetry $\sigma_h$ in the $D_{3h}$ point group, i.e., $\chi^{q,1L}_{\alpha\beta\gamma z}=0$. In the case of normal incidence, the minimum unit structure to hold a quadrupole moment is a bilayer BN that breaks $\sigma_h$. In our transfer matrix computational method, the bilayer quadrupole moment arises from the variation of elementary dipole moments across neighbouring monolayers as a result of the gradient of the pump electric field. For example, consider nonlinear polarization from an AA'-stacked bilayer unit including layer $j$ and $j+1$ is, where the dipole moment vanishes due to centrosymmetry leaving the quadrupole moment as the leading term:
\begin{equation}\label{TMM__BilayerQ}
\begin{aligned}
P_{2\omega,\alpha}^{j,j+1,q} & = P_{2\omega,\alpha}^{j,d} + P_{2\omega,\alpha}^{j+1,d} \\
& = \epsilon_0 \chi^{(2)}_{\alpha\beta\gamma} (E_{\omega, \beta}^{j+1}E_{\omega, \gamma}^{j+1} - E_{\omega, \beta}^{j}E_{\omega, \gamma}^{j}) \\
& = 2 \epsilon_0 \chi^{(2)}_{\alpha\beta\gamma} d_{BN} E_{\omega, \beta}^{j} \frac{ \partial E_{\omega, \gamma}^{j}}{\partial z} + ...
\end{aligned}
\end{equation}
with $d_{BN}$ representing the BN interlayer distance. From ~\eqref{TMM__BilayerQ}, we note that our transfer matrix model is equivalent to taking the quadrupole susceptibility $\chi^{q,2L}_{\alpha\beta\gamma z}$ for an AA'-stacked bilayer unit as:   
\begin{equation}\label{TMM__BilayerQTensor}
\chi^{q,2L}_{\alpha\beta\gamma z} = 2 \chi^{(2)}_{\alpha\beta\gamma} d_{BN} 
\end{equation}
This is consistent with the expectation that relative contribution of increasingly higher-order multipole moments scale like $ak$, where $a$ is the typical atomic dimension of the system, and $k$ is the typical magnitude of wave vector ~\cite{Heinz1991}. 

Next the model computes how the nonlinear polarizations drive second harmonic waves and their propagation throughout the structure. Due to multilayer reflections, the electric field $E_{\omega, \beta}^{j}$ is a superposition of both forward and backward propagating waves:
\begin{equation}\label{TMM_Epump_Forward_Backward}
E_{\omega, \beta}^{j} = E_{\omega, \beta}^{j,+} e^{ik_{\omega}z} + E_{\omega, \beta}^{j,-} e^{-ik_{\omega}z}
\end{equation}
Here $k_\omega=n_\omega k_\omega^0$ is the reduced wavevector, $k_{\omega}^0$ is free-space wavevector, and $n_\omega$ is the linear refractive index all evaluated at the pump frequency $\omega$.

Without loss of generality, the pump polarization is fixed along the $y$ direction, and the twist angle of each layer is $\varphi^{j}$ for layer $j$ as shown in Fig.~\ref{fig:S_TMM}. Therefore, the nonlinear polarization in layer $j$ can be expressed as: 
\begin{equation}\label{TMM_PNL_Y}
P_{2\omega,y}^{j} = \epsilon_0 |\chi^{(2)}| cos(3\varphi^{j}) \left[ E_{\omega,y}^{j,+} e^{ik_{\omega}z} + E_{\omega,y}^{j,-} e^{-ik_{\omega}z} \right]^2
\end{equation}
\begin{equation}\label{TMM_PNL_X}
P_{2\omega,x}^{j} = \epsilon_0 |\chi^{(2)}| sin(3\varphi^{j}) \left[ E_{\omega,x}^{j,+} e^{ik_{\omega}z} + E_{\omega,x}^{j,-} e^{-ik_{\omega}z} \right]^2
\end{equation}
As a result of mixing forward- and backward-propagating waves in Eq. \eqref{TMM_PNL_Y} and Eq. \eqref{TMM_PNL_X}, nonlinear polarizations with different wavevectors ($\pm 2k_\omega$ or $\pm 0$) will be generated. Although the phase change within the source layer $j$ is negligible due to the atomic thickness of monolayer BN, the different wavevectors will lead to appreciably different propagation effects when the whole device structure with hundreds of layers is considered. Therefore the nonlinear polarization pseudovector $\boldsymbol P$ has four contributions accounting for different polarization directions ($x$ or $y$) and different wavevectors ($\pm 2k_\omega$ or $\pm 0$). These are written in the pseudovector form as below:

\begin{equation}\label{TMM_PNL_Pseudovec_y2k}
\boldsymbol P_{2\omega,y}^{j,2k_\omega} = \epsilon_0 |\chi^{(2)}| cos(3\varphi^{j})\begin{bmatrix}
{E_{\omega, y}^{j,+}}^2 \\
{E_{\omega, y}^{j,-}}^2
\end{bmatrix} 
\end{equation}

\begin{equation}\label{TMM_PNL_Pseudovec_x2k}
\boldsymbol P_{2\omega,x}^{j,2k_\omega} = \epsilon_0 |\chi^{(2)}| sin(3\varphi^{j})\begin{bmatrix}
{E_{\omega,y}^{j,+}}^2 \\
{E_{\omega,y}^{j,-}}^2
\end{bmatrix} 
\end{equation}

\begin{equation}\label{TMM_PNL_Pseudovec_y0}
\boldsymbol P_{2\omega,y}^{j,0} = \epsilon_0 |\chi^{(2)}| cos(3\varphi^{j})\begin{bmatrix}
E{_\omega,y}^{j,+}E{_\omega,y}^{j,-} \\
E{_\omega,y}^{j,+}E{_\omega,y}^{j,-}
\end{bmatrix} 
\end{equation}

\begin{equation}\label{TMM_PNL_Pseudovec_x0}
\boldsymbol P_{2\omega,x}^{j,0} = \epsilon_0 |\chi^{(2)}| sin(3\varphi^{j})\begin{bmatrix}
E{_\omega,y}^{j,+}E{_\omega,y}^{j,-} \\
E{_\omega,y}^{j,+}E{_\omega,y}^{j,-}
\end{bmatrix} 
\end{equation}

The nonlinear polarization terms above serve as the sources in generating second harmonic waves. In the low-field regime, depletion of the pump beam can be neglected, and we can solve decoupled inhomogeneous wave equations for $2\omega$ \cite{Shen1984}. As shown by ~\eqref{TMM_PNL_Pseudovec_y2k}-\eqref{TMM_PNL_Pseudovec_x0}, there are in total eight independent nonlinear polarization sources (for two polarizations $x$ and $y$, two wavevectors $2k_\omega$ and $0$, and forward/backward propagating directions). For the y-polarized forward-propagating polarization with a wave vector of $2k_\omega$, the inhomogeneous wave equation in layer $j$ is: 
\begin{equation}\label{TMM_WaveEq_y2k+}
\frac{ \partial^2 E_{2\omega, y}^{j}}{\partial z^2} + k_{2\omega}^2 E_{2\omega, y} = -\frac{4\pi(2\omega)^2}{c^2} P_{2\omega,y}^{j,2k_\omega,+}
\end{equation}
with $k_{2\omega}$ being the reduced wavevector $k_{2\omega}=n(2\omega)2\omega/c$. The solution to~\eqref{TMM_WaveEq_y2k+} is
\begin{equation}\label{TMM_WaveEq_Soln_y2k+}
\begin{aligned}
E_{2\omega, y}^{j,2k_{\omega},+} & = E_{2\omega, y}^{j,+}e^{ik_{2\omega}z} + E_{S, y}^{j,2k_{\omega},+}e^{i2k_{\omega}z} \\
& = E_{2\omega, y}^{j,+}e^{ik_{2\omega}z} + \frac{4\pi}{\epsilon_{\omega}-\epsilon_{2\omega}} P_{2\omega,y}^{j,2k_\omega. +} e^{i2k_{\omega}z}
\end{aligned}
\end{equation}
Here $E_{2\omega, y}^{j,+}$ is the complex amplitude of the homogeneous solution which needs to be further determined by matching boundary conditions in the transfer matrix formalism, and $E_{S, y}^{j,2k_{\omega},+}$ (bound electric field) is the particular solution as determined by the source term. We note that the denominator of the bound electric field explicitly reflects a phase-matching condition, and the divergence is caused by non-depletion assumption of the pump field. Solutions for other polarizations, wavevectors, and propagation directions can be similarly obtained. The resulting particular solutions of bound electric field can be grouped into pseudovector form as:
\begin{equation}\label{TMM_ES_y2k}
\boldsymbol E_{S,y}^{j,2k_\omega} = \frac{4\pi}{\epsilon_\omega-\epsilon_{2\omega}} \boldsymbol P_{2\omega,y}^{j,2k_\omega} 
\end{equation}
\begin{equation}\label{TMM_ES_x2k}
\boldsymbol E_{S,x}^{j,2k_\omega} = \frac{4\pi}{\epsilon_\omega-\epsilon_{2\omega}} \boldsymbol P_{2\omega,x}^{j,2k_\omega} 
\end{equation}
\begin{equation}\label{TMM_ES_y0}
\boldsymbol E_{S,y}^{j,0} = -\frac{4\pi}{\epsilon_{2\omega}} \boldsymbol P_{2\omega,y}^{j,0} 
\end{equation}
\begin{equation}\label{TMM_ES_x0}
\boldsymbol E_{S,x}^{j,0} = -\frac{4\pi}{\epsilon_{2\omega}} \boldsymbol P_{2\omega,x}^{j,0} 
\end{equation}
The bound waves in layer $j$ drives $2\omega$ free waves in adjacent layers ($\boldsymbol E_{2\omega,y}^{i,\pm}$, $\boldsymbol E_{2\omega,x}^{i,\pm}$ in layer $i$, and similarly denoted for layer $k$). The free waves, being solutions to homogeneous wave equations in each layer, have wavevectors of $k_{2\omega}=n_{2\omega}k_{2\omega}^0$. The relationship between the amplitudes of the bound waves in layer $j$ and free waves in the same layer and adjacent layers are determined by matching boundary conditions for the proper type of polarization at $2\omega$. The second harmonic free waves can be solved by using transfer matrix method at $2\omega$~\cite{Bethune1989}. The $y$-polarization components can be derived as below, while the $x$-polarization expressions are similar and omitted. 

\begin{equation}\label{TMM_22}
\boldsymbol E_{2\omega, y}^{k} = \boldsymbol M_{2\omega}^{kj} \boldsymbol \Phi_{2\omega}^{j} (\boldsymbol M_{2\omega}^{ji} \boldsymbol E_{2\omega,y}^{i} + \boldsymbol S_{2\omega,y}^{j,2k_\omega} + \boldsymbol S_{2\omega,y}^{j,0} )
\end{equation}

\begin{equation}\label{TMM_23}
\boldsymbol S_{2\omega,y}^{j,2k_\omega} = (\overline{\boldsymbol \Phi_{2\omega}^{j}} \boldsymbol M_{2\omega}^{js,2k_\omega} \boldsymbol \Phi_{2\omega}^{2k_\omega} - \boldsymbol{M}_{2\omega}^{js,2k_\omega})\boldsymbol{E}_{S,y}^{j,2k_\omega}
\end{equation}

\begin{equation}\label{TMM_24}
\boldsymbol{S}_{2\omega,y}^{j,0} = (\overline{\boldsymbol{\Phi}_{2\omega}^{j}} \boldsymbol{M}_{2\omega}^{js,0} \boldsymbol{\Phi}_{2\omega}^{0} - \boldsymbol{M}_{2\omega}^{js,0})\boldsymbol{E}_{S,y}^{j,0}
\end{equation}

In the equations Eqs. \eqref{TMM_22}-\eqref{TMM_24}, the bound waves in layer $j$ are reformed into the effective source vectors $\boldsymbol{S}_{2\omega,y}^{j,2k_\omega}$ and $\boldsymbol{S}_{2\omega,y}^{j,0}$ for convenience. Note that the matrices $\boldsymbol{M}_{2\omega}^{kj}$ and $\boldsymbol{\Phi}_{2\omega}^{j}$ are ordinary transfer matrices defined for the $2\omega$ frequency free waves, while $\boldsymbol{M}_{2\omega}^{js,2k_\omega}$ and $\boldsymbol{\Phi}_{2\omega}^{2k_\omega}$ are pseudo transfer matrices where Fresnel reflection and transmission coefficients for the second medium (source medium $s$) need to be evaluated using effective indexes $n_S^{j,2k_\omega}$ and $n_S^{j,0}$ corresponding to the bound wave wavevectors $2k_\omega$ and $0$:

\begin{equation}\label{TMM_25}
n_S^{j,2k_\omega} = n_\omega^j, n_S^{j,0}=0
\end{equation}

The $2\omega$ free waves in layer $i$ and $k$ as obtained by Eq. \eqref{TMM_22} can be further propagated through the entire twisted multilayer structure, giving rise to the field of transmitted and reflected SHG: $E_{2\omega,y}^{f,+}(j)$ and $E_{2\omega,y}^{0,-}(j)$. This is obtained by solving linear transfer matrix problem for the multilayer structure at frequency $2\omega$. These are expressed in Eqs.\eqref{TMM_RL1}-\eqref{TMM_RL3} below. Here the index $j$ in the parenthesis denotes that the these output SHG fields are the contribution from nonlinear polarization generated in layer $j$. 

\begin{equation}\label{TMM_RL1}
\boldsymbol{R}_{2\omega}^{jf}\begin{bmatrix} E_{2\omega,y}^{f,+}\\0 \end{bmatrix} - \boldsymbol{L}_{2\omega}^{j0}\begin{bmatrix} 0\\E_{2\omega,y}^{0,-} \end{bmatrix} = (\boldsymbol{S}_{2\omega,y}^{j,2k_\omega} + \boldsymbol{S}_{2\omega,y}^{j,0})
\end{equation}

\begin{equation}\label{TMM_RL2}
\boldsymbol{R}_{2\omega}^{jf} = \overline{\boldsymbol{\Phi}_{2\omega}^j} \boldsymbol{M}_{2\omega}^{jk} \overline{\boldsymbol{\Phi}_{2\omega}^k} \boldsymbol{M}_{2\omega}^{k(k+1)} ... \overline{\boldsymbol{\Phi}_{2\omega}^{f-1}} \boldsymbol{M}_{2\omega}^{(f-1)f}
\end{equation}

\begin{equation}\label{TMM_RL3}
\boldsymbol{L}_{2\omega}^{j0} = \boldsymbol{M}_{2\omega}^{ji} \boldsymbol{\Phi}_{2\omega}^i \boldsymbol{M}_{2\omega}^{i(i-1)} ... \boldsymbol{\Phi}_{2\omega}^2 \boldsymbol{M}_{2\omega}^{21}
\end{equation}

Finally, the reflected SHG field amplitude contributed by each layer are linearly summed up to give the total SHG field polarized along the $y$ direction ($E_{2\omega,y}^{0,-}$) and the $x$ direction ($E_{2\omega,x}^{0,-}$). The corresponding total SHG intensities along $y$ and $x$ directions ($I_{SHG,y}^{tot}$ and $I_{SHG,x}^{tot}$) are computed as squared norm of the coherently summed electric fields:
\begin{equation}\label{TMM_ISHG_tot_par}
I_{SHG,\parallel}^{tot} = I_{SHG,y}^{tot} = {|E_{2\omega,y}^{0,-}|}^2 = \lvert\sum_{j=0}^{f} E_{2\omega,y}^{0,-}(j)\lvert^2 
\end{equation}
\begin{equation}\label{TMM_ISHG_tot_perp}
I_{SHG,\bot}^{tot} = I_{SHG,x}^{tot} = {|E_{2\omega,x}^{0,-}|}^2 = \lvert\sum_{j=0}^{f} E_{2\omega,x}^{0,-}(j)\lvert^2 
\end{equation}
\begin{equation}\label{TMM_ISHG_tot}
I_{SHG}^{tot} = I_{SHG,\parallel}^{tot}  + I_{SHG,\bot}^{tot}
\end{equation}
The contribution from the interface bilayer and the rest of bulk parts of the BN rotator device are then computed as:
\begin{equation}\label{TMM_ISHG_interface}
I_{SHG}^{interface} = \lvert\sum_{j=N_{TOP}}^{N_{TOP}+1} E_{2\omega,y}^{0,-}(j)\lvert^2 + \lvert\sum_{j=N_{TOP}}^{N_{TOP}+1} E_{2\omega,x}^{0,-}(j)\lvert^2 
\end{equation}

\begin{equation}\label{TMM_ISHG_bulk}
\begin{aligned}
I_{SHG}^{bulk} = &\lvert \sum_{j=0}^{N_{TOP}-1} E_{2\omega,y}^{0,-}(j) + \sum_{j=N_{TOP}+2}^{N_{TOP}+N_{BOT}+1} E_{2\omega,y}^{0,-}(j) \lvert^2 \\& + \lvert \sum_{j=0}^{N_{TOP}-1} E_{2\omega,x}^{0,-}(j) + \sum_{j=N_{TOP}+2}^{N_{TOP}+N_{BOT}+1} E_{2\omega,x}^{0,-}(j) \lvert^2  
\end{aligned}
\end{equation}

In this way, we numerically evaluate how interface and bulk SHG contribution varies as a function of twist angle for fabricated micro rotator devices, and compared with experimental data, as shown by Fig.~\ref{fig:S_Bulk_Interface_Contributions}

\subsection{Calculation of interfacial stacking configuration}
\label{SI_GSFE}
\begin{figure*}[ht]
\includegraphics[width=7.2 in]{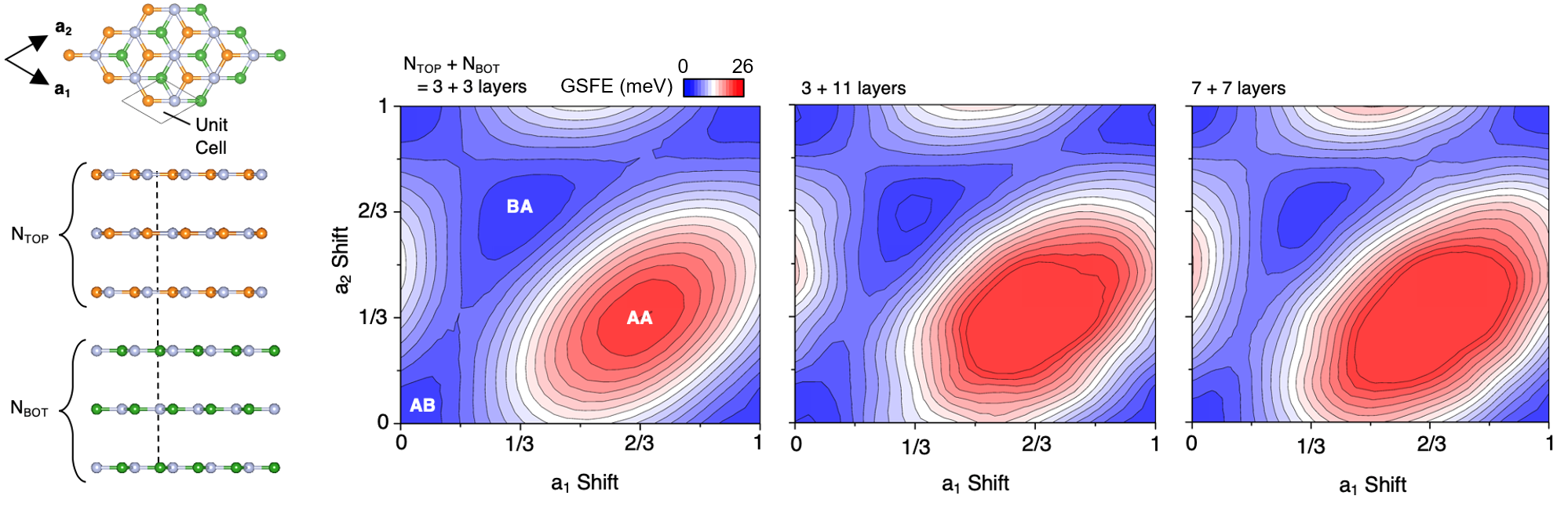}
\caption{\textbf{Schematic atomic structures and energy contours of BN rotators for the twisted angle of 60$^{\circ}$.}
\textbf{a}, Top and side views of the energy minimum of BN rotator with the twisted angle of 60$^{\circ}$.
\textbf{b}, Relative energy contour for the BN rotator with top 3 layers and bottom 3 layers.
\textbf{c} (\textbf{d}), Same quantity as in panel \textbf{b} for the BN rotator with top 7 (3) and bottom 7 (11) layers.
}
\label{fig:S_DFT}
\end{figure*}

We carried out first-principles calculations as implemented in the Vienna Ab initio Simulation Package (VASP)~\cite{Kresse1996}, with Perdew–Burke–Ernzerhof (PBE) exchange-correlation energy functional and projector-augmented wave (PAW) pseudopotentials~\cite{Perdew1996}. We used a plane-wave cutoff of 400 eV and vacuum regions of more than 15 {\AA} between periodically repeated slabs. The first Brillouin zone was sampled using a 10x10x1 k-point grid and vdW interactions were included using the opt88 functional~\cite{Klimes1996}. All structures were fully relaxed until the force on each atom was less than 0.01 eV ${\AA}^{-1}$. Here, we calculated the relative energies of different stacking orders with the AB stacking order ($\Delta E = E_{shift} – E_{AB}$). AB stacking is the lowest-energy configuration for $\theta=60^{circ}$ up to 14 layers as shown by calculation results in Fig.~\ref{fig:S_DFT}, and we reasonably extend that the conclusion holds for bulk BN as well.

\subsection{SHG enhancement factor optimization based on layer numbers}
\label{SI_Gain_Factor_Optimization}
In a homostructure with N$_{TOP}$ and N$_{BOT}$ layers in the top and bottom BN crystal, the SHG intensity I$_{AA'}$ for an AA'-homostructure is dependent on N$_{TOTAL}$ and its respective parity, as shown by Fig.~\ref{fig:1}c in the main text, as well as Fig.~\ref{fig:S_SHG_Ratio_Simulation}c. On the other hand, the SHG intensity I$_{AB}$ for an AB-homostructure is a superposition of the interface dipolar SHG from the bilayer at the interface, and bulk SHG from the N$_{TOP}$+N$_{BOT}$-2 layers left over in the top and bottom bulk BN parts, with simulation results shown by Fig.~\ref{fig:S_SHG_Ratio_Simulation}b. Therefore, the SHG enhancement factor of a single device, defined here as the intensity ratio I$_{AB}$/I$_{AA'}$ between its AB-homostructure form and AA'-homostructure form can be optimized by layer number selections. The simulation results are shown by Fig.~\ref{fig:S_SHG_Ratio_Simulation}a. The results suggest that optimum enhancement factors can be achieved first by selecting N$_{TOTAL}$ to suppress bulk SHG in the AA' interfacial configuration, and secondly selecting N$_{BOT}$ (or equivalently, N$_{TOP}$) to enhance the dipole-allowed SHG at the AB interfacial configuration. Importantly, in the optimized devices (Devices 1 and 2), SHG from the top and bottom bulk BN are both minimized, leaving SHG predominantly coming from the interfacial bilayer, as shown by simluation in Fig.~\ref{fig:S_SHG_Ratio_Simulation}d-f. Finally, experimental and simulation data are compared in Fig.~\ref{fig:S_SHG_Ratio_Simulation}g, showing good agreement.

\begin{figure*}[ht]
\includegraphics[width=7.2 in]{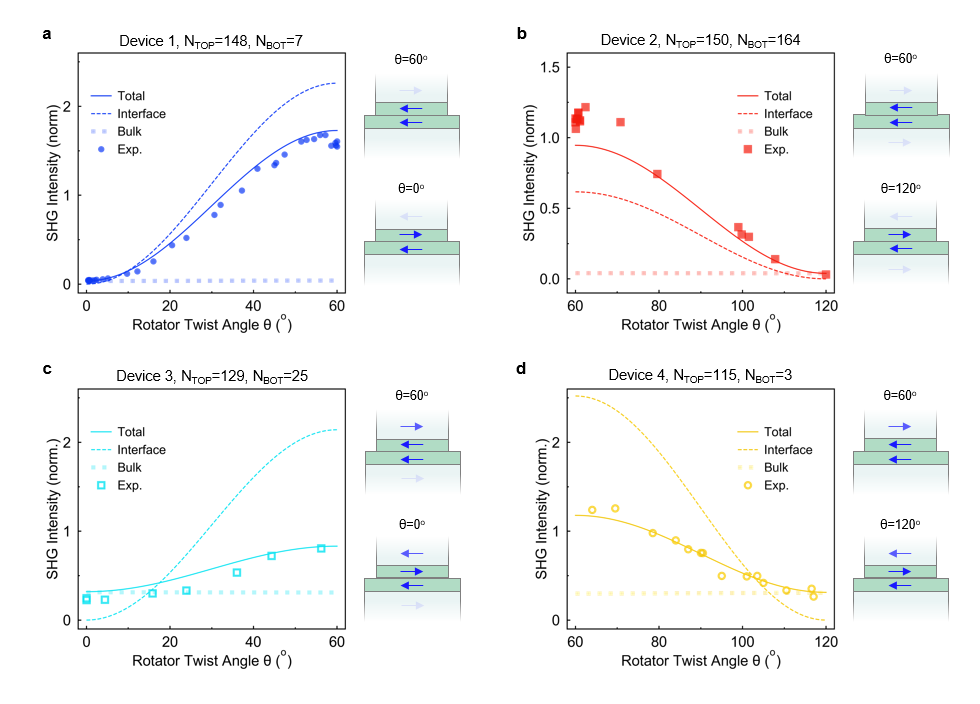}
\caption{\textbf{Bulk and interface SHG contributions in micro rotator devices}
The results are shown for \textbf{a}, device 1, \textbf{b}, device 2, \textbf{c}, device 3, and \textbf{d}, device 4. Markers are experimental data. Solid/dashed/dotted lines are simulation results for total/interface/bulk contributions based on the nonlinear transfer matrix method. The schematics on the right side of each panel illustrate the nonlinear polarization in the interfacial bilayer (pink) and bulk parts (grey). Solid/transparent blue arrows represent high/low magnitude of nonlinear polarization. 
}
\label{fig:S_Bulk_Interface_Contributions}
\end{figure*}

\begin{figure*}[ht]
\includegraphics[width=7.0 in]{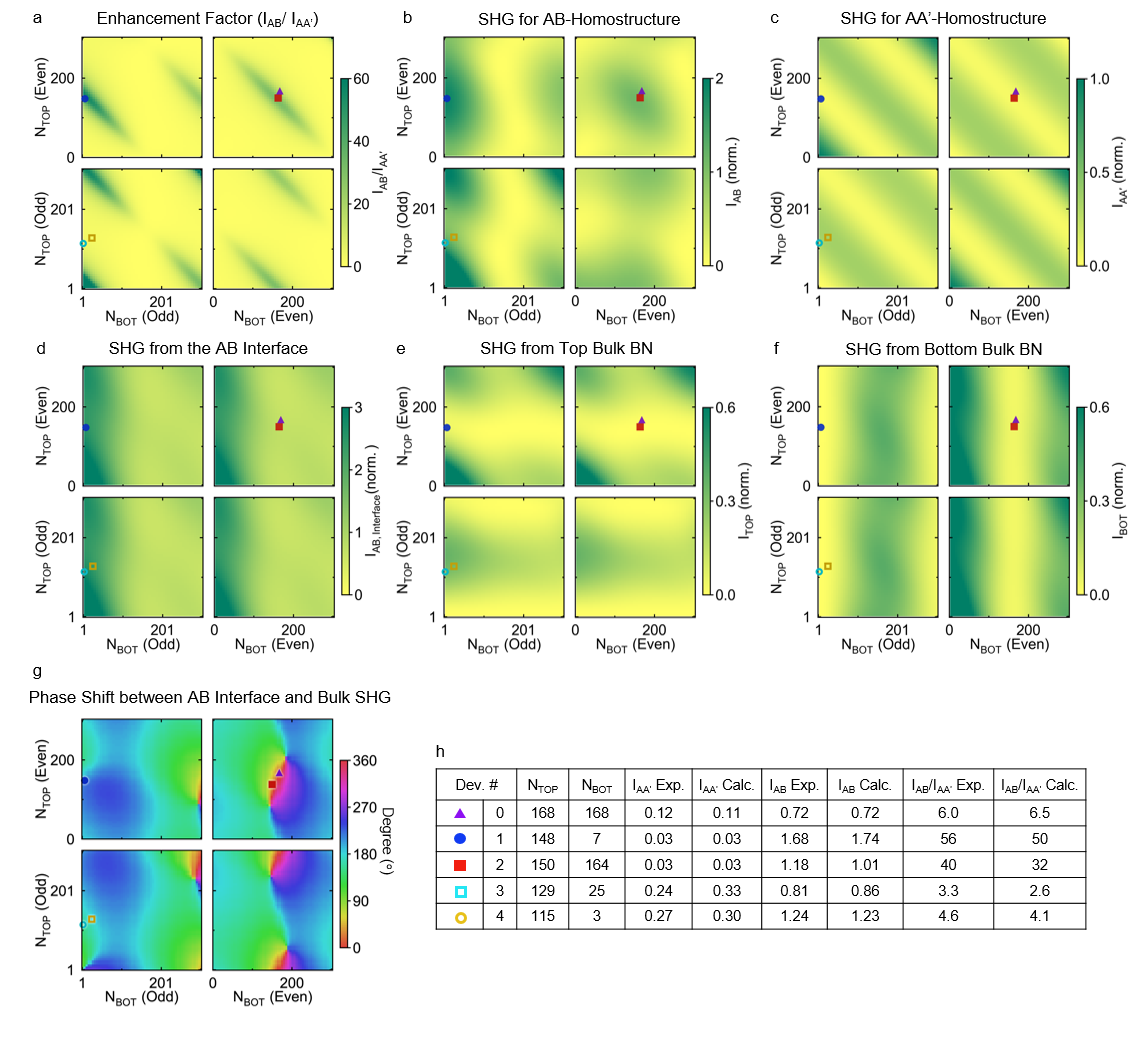}
\caption{\textbf{Simulation results for layer-dependent SHG enhancement factor and phase shift.}
\textbf{a}, Simulation of the SHG enhancement factor I$_{AB}$/I$_{AA'}$ as a function of top and bottom BN layer numbers N$_{TOP}$ and N$_{BOT}$, respectively, calculated by the nonlinear transfer matrix method. I$_{AB}$ and I$_{AA'}$ represents SHG intensity for AB-homostructure and AA'-homostructure interfacial configurations, respectively. The markers tag the layer number configurations of experimentally fabricated devices with SHG results listed in the table shown by \textbf{h}.
Simulation results of the SHG intensity from AB-homostructure I$_{AB}$, AA'-homostructure I$_{AA'}$, AB interfacial bilayer, top bulk BN, and bottom bulk BN are shown by \textbf{b}, \textbf{c}, \textbf{d}, \textbf{e}, \textbf{f}, respectively. Data is normalized to monolayer SHG intensity of BN.
\textbf{g}, The phase shift between reflected SHG from the AB interface and the bulk, as a function of N$_{TOP}$ and N$_{BOT}$.
\textbf{h}, Comparison of experimental and calculated results for all devices.
}
\label{fig:S_SHG_Ratio_Simulation}
\end{figure*}

For micro rotator devices, the twist-angle-dependent experimental and simulation data of SHG intensities are compared  in Fig.~\ref{fig:S_Bulk_Interface_Contributions}. For simulation results, contributions from bulk and interface can be separated. Device 1 features negligible bulk contribution. At $\theta=60^{\circ}$, as shown by the schematics on the right side of Fig.~\ref{fig:S_Bulk_Interface_Contributions}a, the total SHG is dominated by strong interface contribution, with a minor bulk contribution that only slightly counteracts the interface dipole moment. At $\theta=0^{\circ}$, the interface dipole moment vanishes, and the total SHG is close to zero. Similar results are observed for Device 2 which is also interface-dominated, as shown by Fig.~\ref{fig:S_Bulk_Interface_Contributions}b. In Device 3, as shown by the schematics on the right side of Fig.~\ref{fig:S_Bulk_Interface_Contributions}c, the bulk contribution from the top BN is appreciable compared to the interface contribution. This is because the N$_{TOP}$-1 layers in the top BN correspond to a large quadrupole SHG, as shown by Fig.~\ref{fig:1}c. At $\theta=60^{\circ}$, the aligned interface dipole moments are largely counteracted by bulk contribution from the top, resulting in only moderate SHG intensity. At $\theta=0^{\circ}$, despite vanishing interface dipole moment, the bulk contribution still persists, with intensity close to that at $\theta=60^{\circ}$. Therefore the effective tuning range of Device 3 is low. Similar results are observed for Device 4, as shown by Fig.~\ref{fig:1}d.

\subsection{SHG polarization patterns of monolayer and bulk BN}
\label{SI_Bulk_Polarization}
Monolayer BN belongs to the D$_{3h}$ point group ~\cite{Li2013}. The nonvanishing $\boldsymbol \chi^{(2)}$ elements are $\chi^{(2)}_{yyy}=-\chi^{(2)}_{yxx}=-\chi^{(2)}_{xxy}=-\chi^{(2)}_{xyx}$ where $y$($x$) referes to the armchair(zigzag) direction. In our experiments, the pump beam is linearly polarized, and the collected SHG beam with polarization parallel to the pump is selectively detected. In this colinear configuration, the measured SHG intensity $I_{SHG,\parallel}$ will be:
\begin{equation}\label{Eq_monoBN_polarization}
I_{SHG,\parallel} \propto |\chi^{(2)}|^2 cos^2(3\varphi) I_{pump}^2
\end{equation}
where $\varphi$ is the angle between monolayer BN's armchair direction and the pump polarization direction.

\begin{figure*}[ht]
\includegraphics[width=6 in]{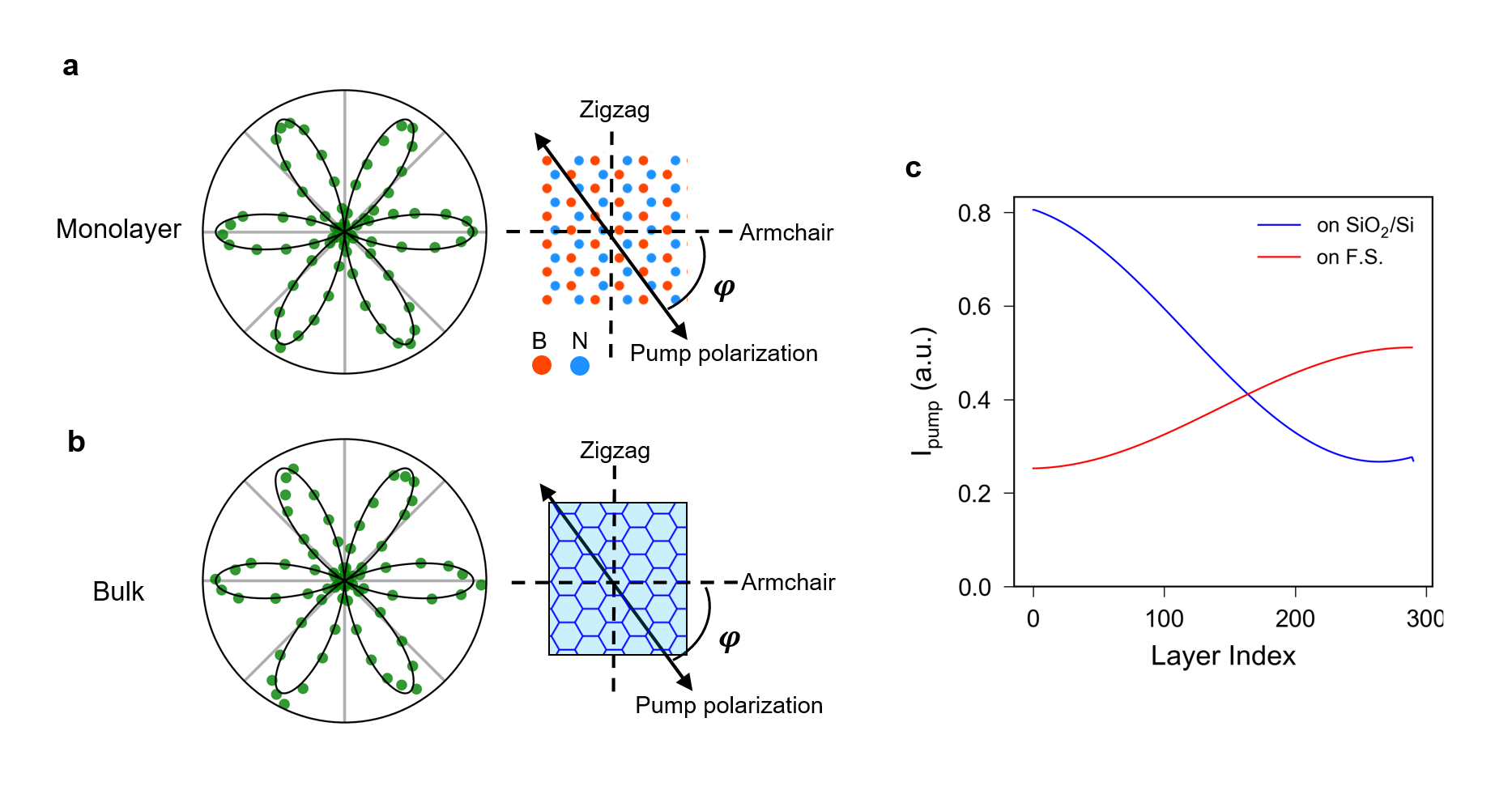} 
\caption{\textbf{Polarization dependent SHG for \textbf{a}, monolayer and \textbf{b}, bulk BN.} Data are measured with the parallel polarization configuration as described in the text. For monolayer BN the sixfold pattern originates from the D$_{3h}$ symmetry. For bulk BN, although the lattice belongs to centrosymmetric D$_{3d}$ point group, inversion symmetry is broken by the gradient of pump field as shown in \textbf{c}, resulting in a net C$_{3v}$ symmetry whose nonvanishing $\boldsymbol \chi^{(2)}$ elements would generate a similar sixfold polarization pattern.
\textbf{c}, The z-directional profile of pump field intensity $I_{pump}$ calculated for a 300-layer BN on a fused silica (F.S.) and a 285-nm-SiO$_2$/Si substrate. The field gradient breaks reflection symmetry.
}
\label{fig:S_Bulk_Polarization}
\end{figure*}

Bulk hexagonal BN crystal belongs to the centrosymmetric D$_{3d}$ point group with preserved inversion symmetry, as a result of the AA' stacking between each neighbouring layers. However, when such a bulk BN film is placed on a substrate with finite refractive index, the pump field may become highly nonuniform with a large gradient along the z-direction. This is illustrated by Fig.~\ref{fig:S_Bulk_Polarization}c which shows the calculated $z$-directional profile of pump field intensity $I_{pump}$ inside a 300-layer BN on top of a fused silica (F.S.) substrate and a 285-nm-SiO$_2$/Si substrate. The gradient of pump field breaks reflection symmetry along the x-y plane (\emph{i.e.}, the plane including armchair and zigzag orientations). Therefore, dressed by the optical field, the net symmetry is reduced from D$_{3d}$ to C$_{3v}$ which has nonvanishing $\boldsymbol \chi^{(2)}$ elements of $\chi^{(2)}_{yyy}=-\chi^{(2)}_{yxx}=-\chi^{(2)}_{xxy}=-\chi^{(2)}_{xyx}$, $\chi^{(2)}_{xzx}=\chi^{(2)}_{yzy}$, $\chi^{(2)}_{xxz}=\chi^{(2)}_{yyz}$, $\chi^{(2)}_{zxx}=\chi^{(2)}_{zyy}$, and $\chi^{(2)}_{zzz}$. As a result, the same in-plane polarization dependence as in Eq. ~\eqref{Eq_monoBN_polarization} would be expected, \emph{i.e.}, a sixfold flower pattern, where the maximum intensity is aligned with armchair orientation as shown by ~\ref{fig:S_Bulk_Polarization}b. 

\subsection{Locking behavior of BN rotators}
\label{SI_Locking}
\begin{figure*}[ht]
\includegraphics[width=7.2 in]{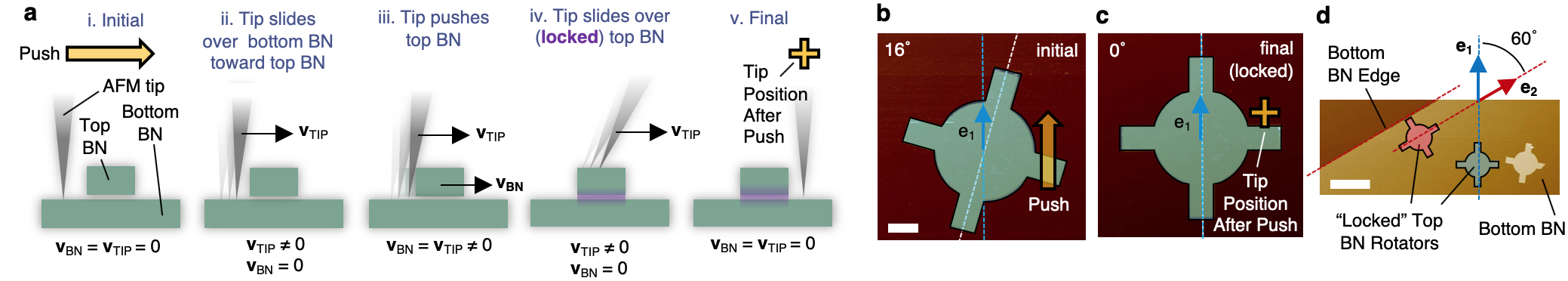} 
\caption{\textbf{Locking behavior observed for a vdW homojunction}
\textbf{a}, Push progression through a locking state, with velocity of the top BN rotator arm, $v_{BN}$, and the velocity of the tip, $v_{TIP}$, indicated for states i-v with labels and black arrows when the velocity is not equal to zero. Yellow arrow represents the planned push, and the yellow plus-symbol represents the tip position after the push. Purple highlight added to iv-v. to indicate a locked-interface.
AFM scans before \textbf{b}, and after \textbf{c} pushing the top BN crystal with an AFM tip. The rotator axis is illustrated with a white-dashed line, and the axis of a lattice unit vector ($e_1$, blue arrow) is illustrated with a blue-dashed line. Scale bar for \textbf{b-c} is 1 $\mu$m.
\textbf{d}, The locking axes of rotators are always 60$^{\circ}$ apart, as illustrated with lattice vectors \textbf{$e_1$} (blue) and \textbf{$e_2$} (red) parallel to the rotation axes of the two locked rotators, highlighted in blue and red. The rotator highlighted in light-blue in \textbf{d} is the same rotator shown \textbf{c} in its final state. The scale bar is 5 $\mu$m.
}
\label{fig:S_locking}
\end{figure*}

In all rotatable homostructures of BN we observe locking at angles set 60$^{\circ}$ apart, meaning that at these special rotation states the BN rotators can no longer be pushed by the AFM tip. Fig.~\ref{fig:S_locking}a illustrates the stages of a push sequence during which the locking occurs, and Fig.~\ref{fig:S_locking}b-d shows AFM scans before and after the progression illustrated in Fig.~\ref{fig:S_locking}a. The locking condition is found to perfectly correspond with crystallographic directions as corroborated both by alignment with straight BN edges and SHG characterization, and can therefore serve as a reference for intermediate rotation angles as measured from AFM scans.

\subsection{BN-Graphene-BN devices: effect of interface modification on SHG}
\label{SI_BNGBN}
\begin{figure*}[ht]
\includegraphics[width=7 in]{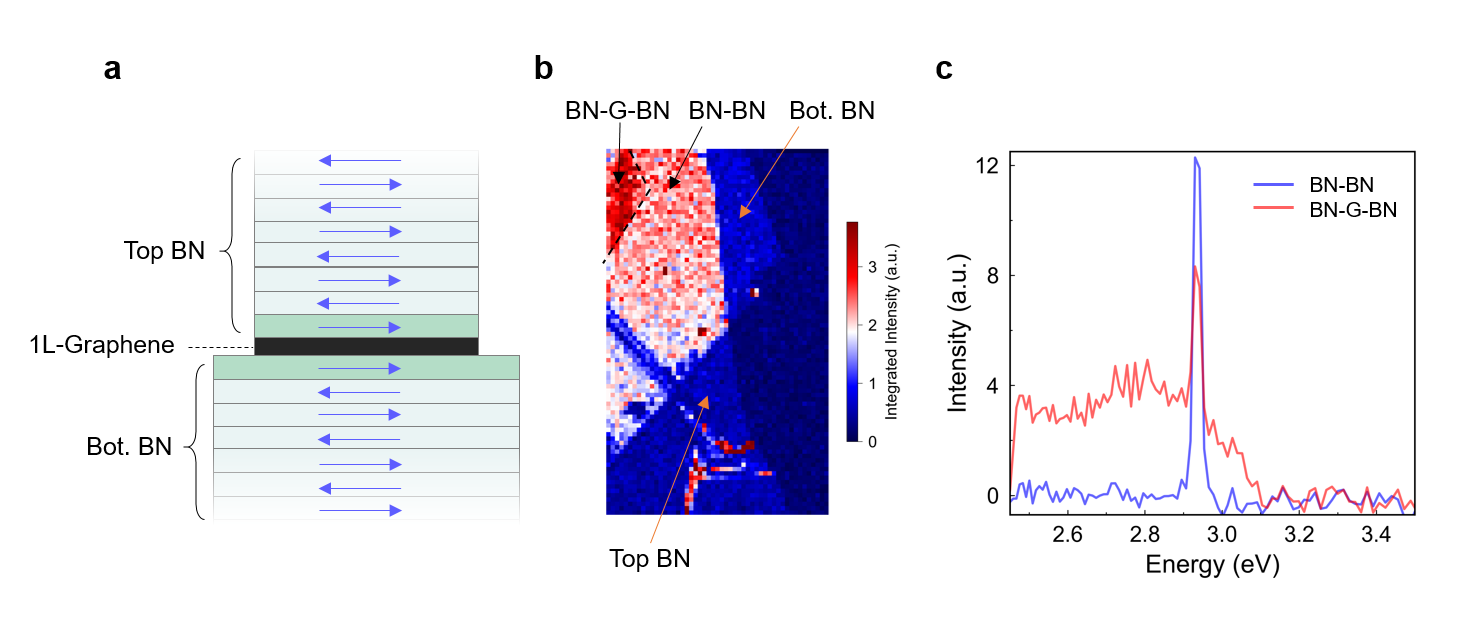} 
\caption{\textbf{SHG from BN-graphene-BN heterostructure.}
\textbf{a}, Cross-sectional schematics of the BN-graphene-BN device. The top BN and bottom BN form AB-homostructure as in Fig.~\ref{fig:1}d. The monolayer (1L) graphene is aligned with both top and bottom BN.
\textbf{b}, Confocal scanning image of the sample using 820 nm pulsed laser. Regions of BN-G-BN, BN-BN, and top and bottom BN crystals are labelled.
\textbf{c}, Emission spectra collected from the BN-G-BN region (red) and BN-BN region (blue). The BN-BN region shows a prominent SHG peak, while the BN-G-BN region shows much weaker SHG with a broadband fluorescence signal (likely from the graphene).
}
\label{fig:S_BNGBN}
\end{figure*}

In order to further consolidate verify the dominant role of the BN-BN interface in generating the SHG signal, monolayer graphene (G) is inserted encapsulated in between the top and bottom BN of an AB-homostructure, forming a stationery BN-G-BN device, as shown by the schematic in Fig.~\ref{fig:S_BNGBN}a. The graphene lattice is aligned with both top and bottom BN. Owing to its centrosymmetric crystal structure, monolayer graphene itself does not generate additional SHG under normal in-plane excitation. However, the inserted encapsulated graphene layer may substantially alter interlayer charge transfer between the interfacial bilayer BN (shaded green), and subsequently change the oscillator strength responsible for interfacial dipolar SHG. On the other hand, due to its monolayer thickness, the extra layer of graphene will only induce negligible optical phase shift to affect any bulk SHG contribution. Therefore, we can further verify the interface-dominated SHG by observing whether an inserted encapsulated monolayer graphene can cause appreciable changes in SHG intensity.

Fig.~\ref{fig:S_BNGBN}b shows a confocal image of a device with a BN-G-BN region, a BN-BN AB-homostructure region, as well as bare bottom and top BN regions. The emission spectra from the BN-G-BN and BN-BN regions are compared in Fig.~\ref{fig:S_BNGBN}c. Note that we observed a broadband of fluorescence signal from graphene, which makes the BN-G-BN appear brighter in the image in Fig.~\ref{fig:S_BNGBN}b. For the narrow-band SHG intensity, we found it to be significantly reduced in the BN-G-BN region as compared to BN-BN AB-homostructure region, validating our assignment of the SHG as dominated by the interfacial bilayer rather than the bulk. We note that similar results are also reported in Ref.~\cite{Stepanov2020}.

\subsection{Extended SHG polarization data from BN micro rotators: effect of substrate and bulk contributions}
\label{SI_Extended_Polarization_Data}
In the devices designed for maximizing the interface SHG contribution, the polarization patterns are determined by the D$_{3}$ symmetry of the moir\'e bilayer lattice and they closely follow the rotation of moir\'e angle. On the other hand, in devices with strong bulk SHG response, the polarization pattern reflects the reduced C$_3$ symmetry.

\begin{figure*}[ht]
\includegraphics[width=6 in]{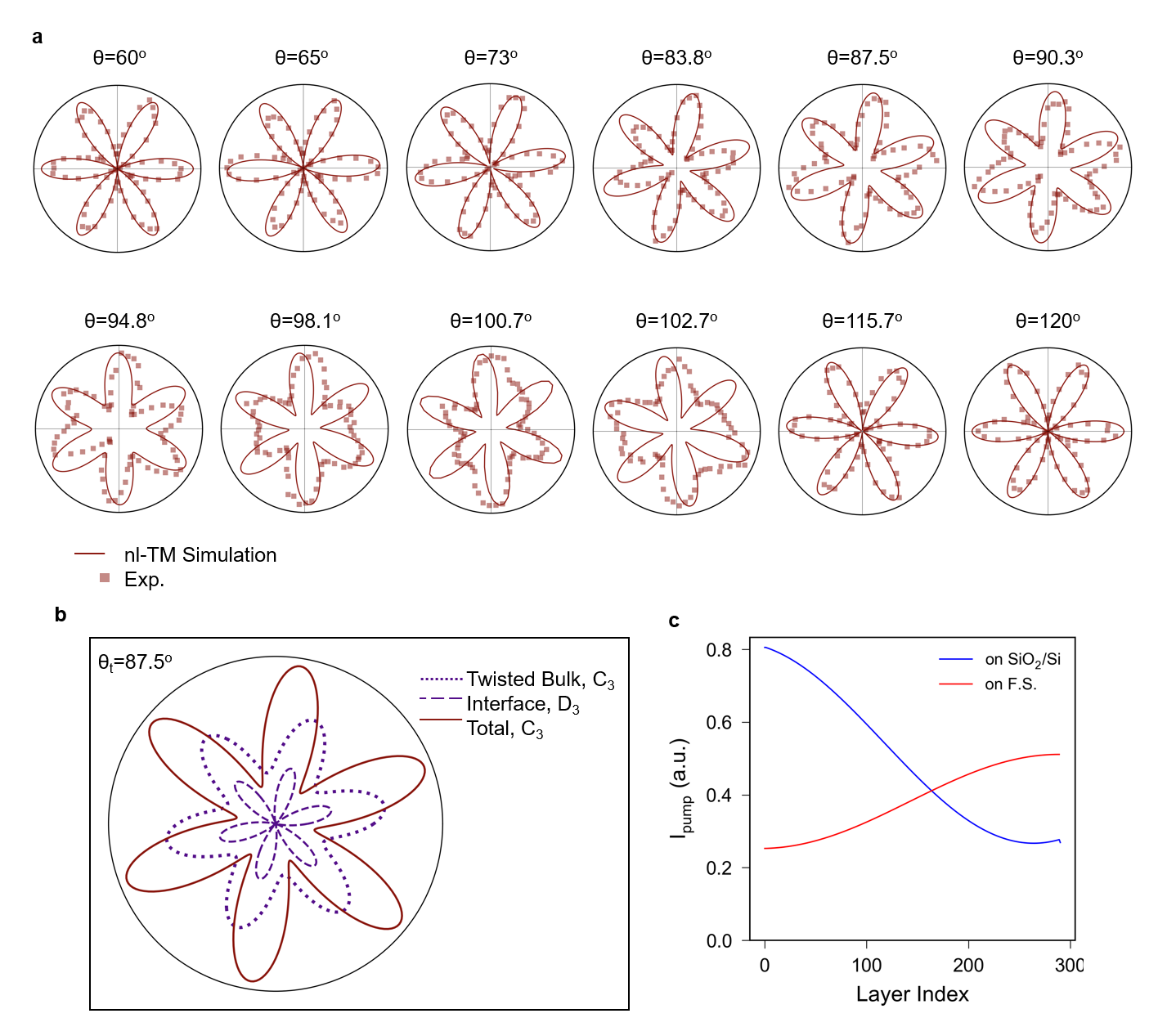} 
\caption{\textbf{SHG polarization pattern measured for device 3 on 285 nm  SiO$_2$/Si substrate, showing competing interface and bulk effects.} \textbf{a}, Extended polarization data at various twist angles. Red squares are experimental data and solid lines are simulation results of this device based on nonlinear transfer matrix method (nl-TM). \textbf{b}, Simulated polarization patterns for contribution from the interface (blue dashed line), the bulk(blue dotted line), and the total device (red solid line). The interface bilayer belongs to $D_3$ point group, and the non-vanishig $\chi^{(2)}$ elements are $\chi^{(2)}_{xxx}=-\chi^{(2)}_{xyy}=-\chi^{(2)}_{yxy}=-\chi^{(2)}_{yyx}$. The resulting polarization pattern is a sixfold flower (without any isotropic center parts, \emph{i.e.}, nodal points reaching zero) and maximum intensity angle aligned along the $C2 (x)$ axes. The twisted bulk structure has a reduced $C_3$ symmetry, and has two independent sets of non-vanishig $\chi^{(2)}$ elements: $\chi^{(2)}_{xxx}=-\chi^{(2)}_{xyy}=-\chi^{(2)}_{yxy}=-\chi^{(2)}_{yyx}$, $\chi^{(2)}_{yyy}=-\chi^{(2)}_{yxx}=-\chi^{(2)}_{xxy}=-\chi^{(2)}_{xyx}$. The resultant polarization pattern will not necessarily reach zero at the nodal points and therefore shows an isotropic center pattern. (c) Simulated profile of pump field intensity through the BN layers. When on 285 nm SiO$_2$/Si substrate, interference effects result in a larger $z$-gradient that breaks centrosymmetry, leading to stronger bulk SHG contribution.
}
\label{fig:S_Polarization_Dev3_SiO2}
\end{figure*}

\begin{figure*}[ht]
\includegraphics[width=6 in]{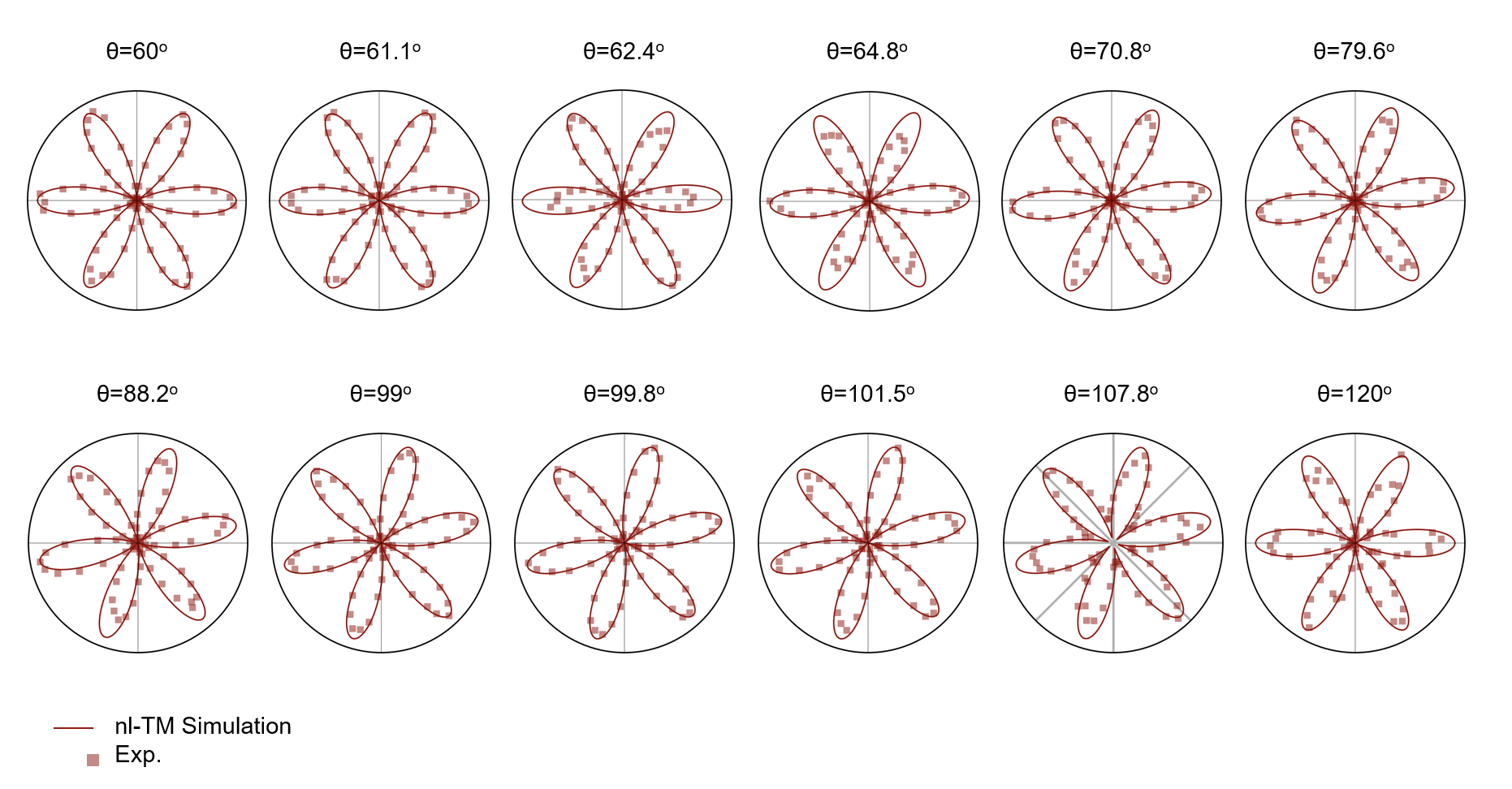} 
\caption{\textbf{SHG polarization patterns measured for device 3 transferred onto a fused silica substrate.} Red squares are experimental data and solid lines are simulation results from the nonlinear transfer matrix calculation. 
}
\label{fig:S_Polarization_Dev3_FS}
\end{figure*}

\begin{figure*}[ht]
\includegraphics[width=6 in]{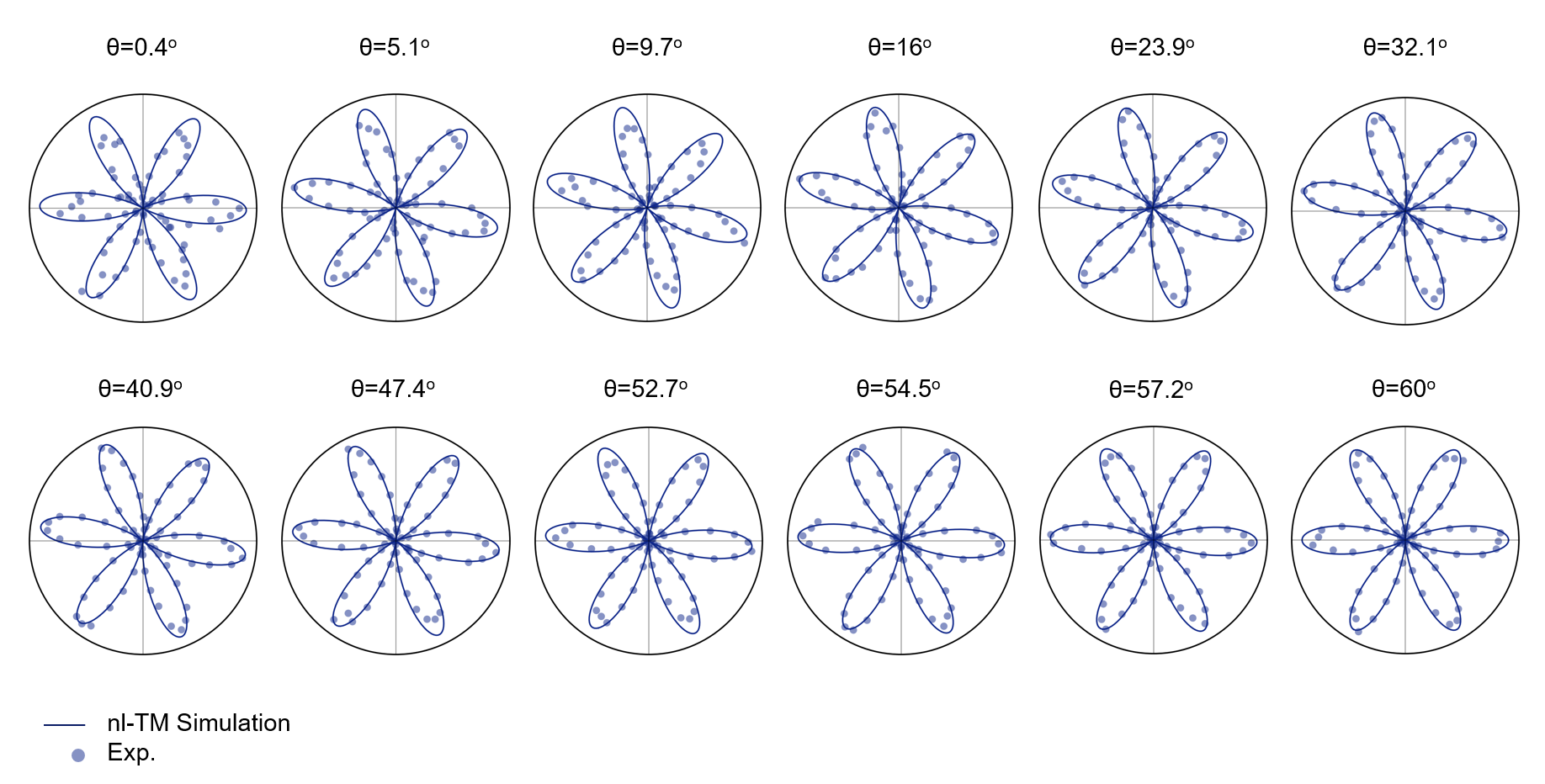} 
\caption{\textbf{SHG polarization patterns measured for device 1 on a fused silica substrate.} Blue dots are experimental data and solid lines are simulation results from the nonlinear transfer matrix calculation. 
}
\label{fig:S_Polarization_Dev1_FS}
\end{figure*}

Fig.~\ref{fig:S_Polarization_Dev3_SiO2}a shows the polarization patterns measured for Device 2 (N$_{TOP}$=150, N$_{BOT}$=138) on 285 nm SiO$_2$/Si substrate. Compared to polarization patterns of interface-dominated devices (see Fig.~\ref{fig:S_Polarization_Dev3_FS} and Fig.~\ref{fig:S_Polarization_Dev1_FS}), bulk-dominated devices show appreciable nonzero isotropic center circles. As shown in Fig.~\ref{fig:S_Bulk_Polarization}c, an interference effect from the presence of the 285 nm thin film of SiO$_2$ on top of the Si substrate induces a significant breaking of reflection symmetry with respect to the interface plane, and thus causes strong bulk SHG contribution in this device when placed on the 285 nm SiO$_2$/Si substrate. With arbitrary top-bottom twist angle $\theta$, and a broken reflection symmetry, the net symmetry has to be further reduced to only C$_3$. The in-plane nonvanishing elements for C$_3$ are $\chi^{(2)}_{xxx}=-\chi^{(2)}_{xyy}=-\chi^{(2)}_{yxy}=-\chi^{(2)}_{yyx}$, $\chi^{(2)}_{yyy}=-\chi^{(2)}_{yxx}=-\chi^{(2)}_{xxy}=-\chi^{(2)}_{xyx}$. The resulting expression for parallel-polarized SHG intensity $I_{SHG,\parallel}$ is:
\begin{equation}\label{Eq_C3_polarization}
I_{SHG,\parallel} \propto [|\chi^{(2)}_{yyy}|cos(3\varphi) +  |\chi^{(2)}_{xxx}| sin(3\varphi)]^2I_{pump}^2
\end{equation}
As shown by Eq. ~\eqref{Eq_C3_polarization}, the resulting polarization pattern may not have zero nodal points. Fig.~\ref{fig:S_Polarization_Dev3_SiO2}b shows the simulated polarization pattern from bulk and interface contributions, for a twist angle of $87.5^\circ$ for the structure of Device 2 (on 285 nm SiO$_2$/Si substrate). The simulation is performed by the nonlinear transfer matrix method. The interface polarization shows a sixfold flower pattern with zero nodal points, while the bulk polarization has a nonzero center circle which is characteristic of the C$_3$ (as in Eq.~\eqref{Eq_C3_polarization}). With the combination of bulk and interface effects, the total SHG polarization also shows C$_3$ symmetry.

In comparison, Fig.~\ref{fig:S_Polarization_Dev3_FS} shows the polarization patterns for \emph{the same device} (Dev. 2, N$_{TOP}$=150, N$_{BOT}$=138) when it's transferred onto a bulk fused silica substrate. On a substrate such as this, with no oxide thin film below the bottom BN, the gradient of pump field is much smaller and there's significantly less SHG contribution from bulk BN (Fig.~\ref{fig:S_Polarization_Dev3_SiO2}). In addition, Fig.~\ref{fig:S_Polarization_Dev1_FS} shows polarization patterns measured for Device 1 (N$_{TOP}$=7, N$_{BOT}$=148) on fused silica substrate, which is also optimally designed for maximizing the interface effect. The polarization patterns for both are sixfold flowers (with negligible isotropic centers) generated by the D$_3$ interface moir\'e superlattice.

\subsection{Fabrication of stationary devices with AB and AA' stacked interfaces}
\label{SI_Fab_Stationery_Devices}
The stationary device shown in Fig.~\ref{fig:1}e in the main text, Device 0, includes both AB and AA' stacked interfaces. In order to achieve this configuration, a bulk BN crystal with uniform thickness is broken into multiple pieces during the exfoliation process, and re-stacked with 0$^{\circ}$ and 60$^{\circ}$ of relative rotation using a high-precision rotation stage in the transfer station setup~\cite{Kim2016}. Fig.~\ref{fig:S_dev0_fab}a shows the fractured BN crystal used for assembling Device 0. Fig.~\ref{fig:S_dev0_fab}b-h shows the pick-up sequence for the bulk BN crystal of thickness $N=168$ layers, with regions of thickness $2N=336$ layers that include an artificially stacked interface between the BN layers. The PPC film is first touched-down and spread over the surface of the target fractured piece, as shown in Fig.~\ref{fig:S_dev0_fab}b. After each pick-up shown in Figs.~\ref{fig:S_dev0_fab}c-h, the pieces of BN mounted on the surface of the PPC film above the substrate surface appear less saturated in color than the BN still mounted on the SiO$_2$ surface, and interfacial homostructure regions formed by overlapping BN crystals appear darker in color. At an intermediate point during assembly, the remaining target BN crystal pieces on the substrate are rotated 60$^{\circ}$ relative to the BN crystals mounted on the PPC film, as shown in Fig.~\ref{fig:S_dev0_fab}e. Fig.~\ref{fig:S_dev0_fab}i shows the final twist configuration of Device 0. We observe minimal sliding of the interfaces during transfer to a fused silica substrate, and after annealing, presumably because the crystallographically aligned interfaces are at their lowest energy stacking configurations and are therefore locked in place.

This technique is adapted from those described in the supplementary information of Ref.~\cite{Finney2019} for fabrication of stationary double-aligned graphene-BN heterostructures, and can be extended to deterministically set a collection of alignment states for any broken vdW crystal with uniform layer number thickness.

\subsection{Mechanical relaxation near 60$^\circ$ twist}
\label{SI_relaxation}
\begin{figure*}[ht]
\includegraphics[width=7 in]{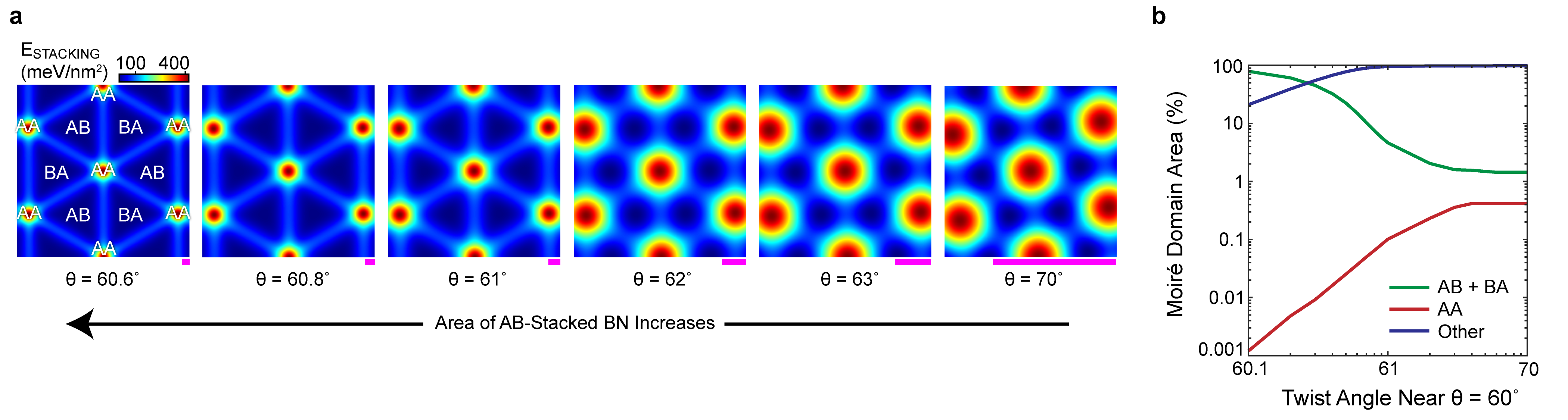}
\caption{\textbf{Mechanical relaxation of twisted bilayer BN near $\theta=60^{\circ}$}
\textbf{a}, Stacking fault energy density, $E_{STACKING}$, for twisted bilayer BN moir\'e unit cell as a function of twist angle near $\theta=60^{\circ}$. Domains identified by their corresponding stacking order are labeled on the leftmost plot. Direction of increasing AB/BA domain area indicated below. Magenta scale-bars at the lower right of each map are all 1 nm.
\textbf{b}, Moir\'e domain area as a function of twist angle near $\theta=60^{\circ}$. The total percent area of energetically favorable AB/BA stacking is plotted in green, energetically unfavorable AA in red, and all other stacking in blue. AB/BA (AA) stacking energy is defined as being within 5 meV/nm$^2$ of the minimum (maximum) $E_{STACKING}$ computed.
}
\label{fig:S_relax}
\end{figure*}

\begin{figure*}[ht]
\includegraphics[width=7.2 in]{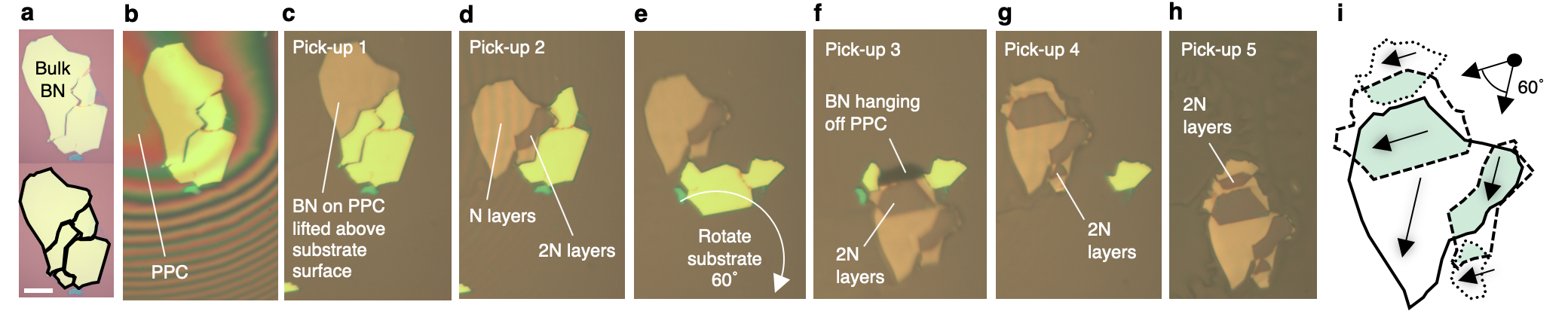}
\caption{\textbf{Assembly of a BN homostructure with both AB and AA' stacked interfaces.}
\textbf{a}, Bulk BN crystal on SiO$_2$ (top) with black outline (bottom) indicating fracture lines. Scale bar is 10 $\mu$m
\textbf{b}, Spreading of the poly-propylene carbonate (PPC) film over the BN crystal prior to pick-up 1. Touched-down region labeled as PPC, thin film interference pattern formed by the thin layer of air between the PPC and the target substrate.
\textbf{c-h}, Stack in-progress immediately following each pick-up, with an intermediate rotation of the substrate by 60$^{\circ}$ shown in \textbf{e}.
\textbf{i}, Final relative twist configuration of the BN pieces for device 0 discussed in the main text, with overlap regions highlighted in light-green.
}
\label{fig:S_dev0_fab}
\end{figure*}

Atomic relaxation calculations of twisted bilayer BN near $\theta=60^\circ$ were performed following the method presented in Ref.~\cite{Carr2018}. The function parameters of the generalized stacking fault energy were taken from Ref.~\cite{Zhou2015}, and the elastic coefficients of single BN layer were taken from Ref.~\cite{Falin2017}. The total energy, composed of a stacking energy term and an elastic energy term, was minimized within a continuous model to produce the local displacement fields of the relaxed system. Fig.~\ref{fig:S_relax}a shows the stacking energy density, $E_{STACKING}$, plotted in real space over the moir\'e unit cell for several twist angles near $\theta=60^\circ$. It is qualitatively apparent that as the twist angle approaches $\theta=60^\circ$ the area of mechanically relaxed BN stacked in the energetically favorable AB or BA stacking increases.

We illustrate this onset of mechanical relaxation quantitatively in Fig.~\ref{fig:S_relax}b, where we show a rapid increase in the percent-area taken up by AB and BA stacked BN within the moir\'e unit cell, and a rapid decrease in the percent-area taken up by AA stacked BN within the moir\'e unit cell as the twist angle decreases past $61^\circ$. The range in values of $E_{STACKING}$ assigned to the energetically degenerate AB and BA stacking is taken to be within 5 meV/nm$^2$ of the minimum $E_{STACKING}$ computed, and similarly the range in values of $E_{STACKING}$ assigned to AA stacking is arbitrarily taken to be within 5 meV/nm$^2$ of the maximum $E_{STACKING}$ computed. Any value of $E_{STACKING}$ falling outside of these ranges is labeled as \textit{other}, a designation that dominates at large angle since the twisted BN lattices are incommensurate in this regime.

The symmetry pinning effect discussed in the main text and shown in Fig.~\ref{fig:3}f is potentially related to the rapid increase in the areal coverage of symmetry-broken AB and BA stacked BN at a twist angle near $61^\circ$. The bottom BN is much larger than the rotator, and presumably well-coupled to the fused silica substrate, mechanically. If the bottom BN can be considered microscopically stationary with respect to the lab frame, then in the presence of relaxation effects the top BN should locally twist to match the rigid bottom BN within the AB and BA domains. In this scenario the AB and BA domains would be effectively aligned to the bottom BN crystal, even though the moir\'e pattern itself would be at some angle $\theta_M$. This would generate SHG inconsistent with what we predict for the non-zero global twist of the top BN rotator. This hypothesis is difficult to test without direct visualization of the buried moir\'e. Future experiments involving scanning probe measurements and polarization resolved SHG mapping with resolution below the scale of the moir\'e pattern features would resolve the microscopic symmetries and would detail the effects of mechanical relaxation. Advancements in experimental techniques combined with the modeling techniques adopted here are likely necessary to understand this symmetry pinning effect.



\begin{figure*}[ht]
\includegraphics[width=6.0 in]{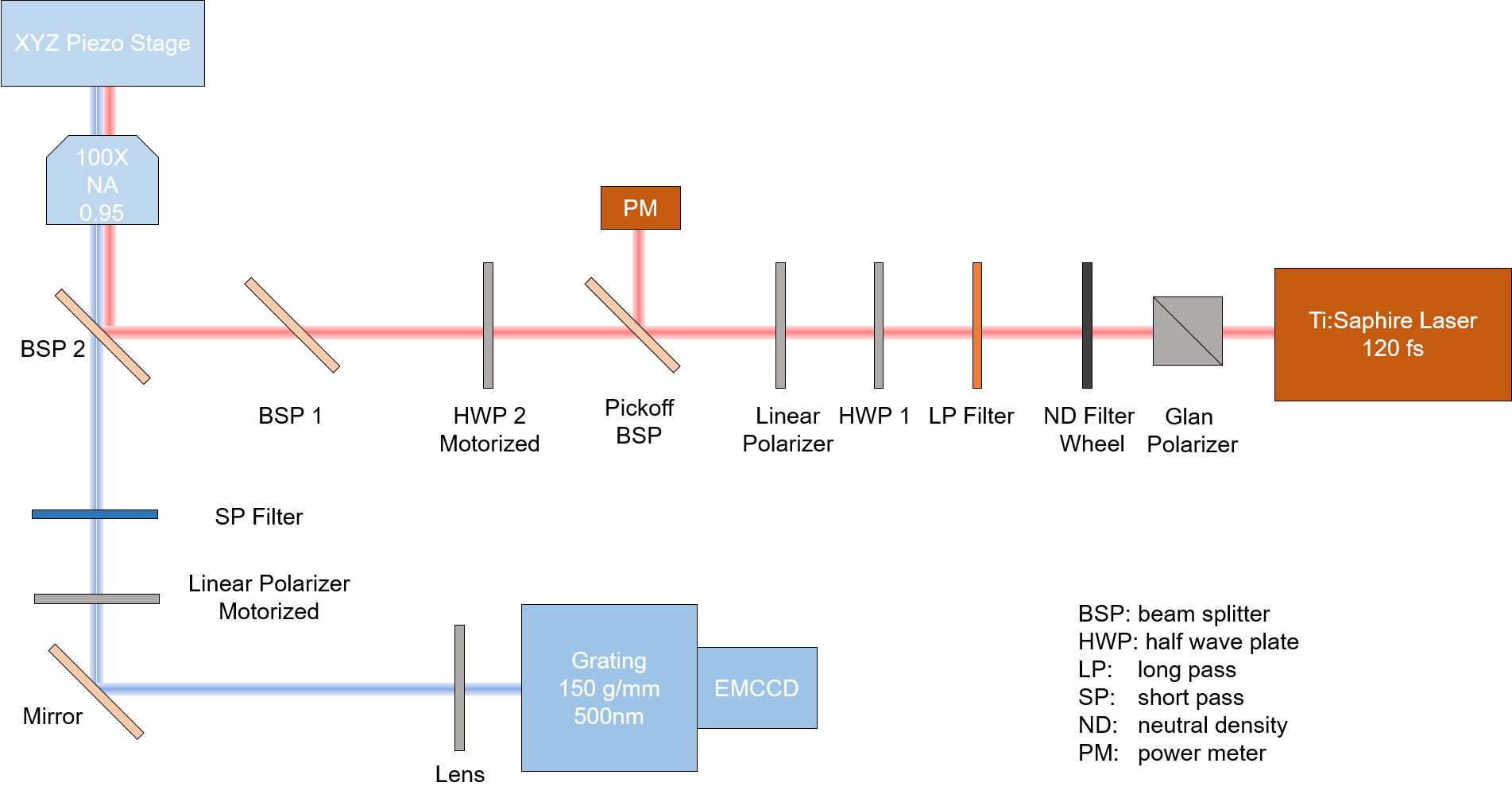} 
\caption{\textbf{Experimental setup for SHG measurements.}
}
\label{fig:S_setup}
\end{figure*}

\end{document}